\documentclass[12pt]{article}
\pdfoutput=1
\usepackage{jheppub, amsmath,amssymb,amsfonts,amsxtra, mathrsfs, makeidx,graphics,graphicx,amsthm,epsfig, ytableau,bm,longtable,float, color,tikz,mathtools,xfrac}
\usetikzlibrary{positioning}

\usepackage{bbm}

\newcommand{\eg}{\textit{e.g.}}

\newcommand{\ie}{\textit{i.e.}}

\numberwithin{equation}{section}

\newcommand{\nn}{\nonumber}

\newcommand{\be}{\begin{equation}} \newcommand{\ee}{\end{equation}}
\newcommand{\bea}{\begin{equation} \begin{aligned}} \newcommand{\eea}{\end{aligned} \end{equation}}

\def\tilde{\widetilde}

\def\hat{\widehat}

\def\bar{\overline}

\def\rt2{\sqrt{2}}

\def\CH{{\cal H}}

\def\CN{{\cal N}}

\def\1{{\ds 1}}

\newcommand{\cA}{\mathcal{A}}

\newcommand{\cK}{\mathcal{K}}

\newcommand{\cN}{\mathcal{N}}

\newcommand{\bC}{\mathbb{C}}

\newcommand{\bZ}{\mathbb{Z}}

\def\repa{\raise4pt\hbox{$\square$}\mkern-14mu\raise-4pt\hbox{$\square$}}
\def\repab{\overline{\raise4pt\hbox{$\square$}\mkern-14mu\raise-4pt\hbox{$\square$}\mkern-1mu}}

\def\smileface{\ensuremath{\hbox{\large$\bigcirc$}\mkern-15mu\raise-1pt\hbox{\scriptsize$\smallsmile$}%
\mkern-10mu\raise4pt\hbox{..}\mkern4mu}}
\def\frownface{\ensuremath{\hbox{\large$\bigcirc$}\mkern-15mu\raise-1pt\hbox{\scriptsize$\smallfrown$}%
\mkern-10mu\raise4pt\hbox{..}\mkern4mu}}

\def\node#1#2{\overset{#1}{\underset{#2}{{\color{gray} \bullet}}}}
\def\Node#1#2{\overset{#1}{\underset{#2}{{ \bullet}}}}
\def\ver#1#2{\overset{{\llap{$\scriptstyle#1$}\displaystyle{\color{gray} \bullet}{\rlap{$\scriptstyle#2$}}}}{\scriptstyle\vert}}
\def\Ver#1#2{\overset{{\llap{$\scriptstyle#1$}\displaystyle\bullet{\rlap{$\scriptstyle#2$}}}}{\scriptstyle\vert}}

\def\Uer#1#2{\underset{{\llap{$\scriptstyle#1$}\displaystyle\bullet{\rlap{$\scriptstyle#2$}}}}{\scriptstyle\vert}}
\usetikzlibrary{arrows}
\tikzstyle{every picture}+=[remember picture]
\tikzstyle{na} = [baseline=-.5ex]

\newcommand{\ba}{\begin{array}}
\newcommand{\ea}{\end{array}}
\newcommand{\bi}{\begin{itemize}}
\newcommand{\ei}{\end{itemize}}
\def\vec#1{\bm{#1}}
\def\bea#1\eea{\allowdisplaybreaks \begin{align}#1\end{align}}
 \newcommand{\ben}{\begin{enumerate}}
\newcommand{\een}{\end{enumerate}}
\newcommand{\bean}{\begin{eqnarray*}}
\newcommand{\eean}{\end{eqnarray*}}
\newcommand{\eref}[1]{(\ref{#1})}

\newcommand{\tref}[1]{Table~\ref{#1}}
\newcommand{\fref}[1]{Figure~\ref{#1}}
\newcommand{\PE}{\mathop{\rm PE}}

\newcommand{\BC}{\mathbb{C}}

\newcommand{\BZ}{\mathbb{Z}}

\newcommand{\comment}[1]{}

\newcommand{\fflat}{\mathcal{F}^\flat}

\title{Coulomb branch Hilbert series \\ and Three Dimensional Sicilian Theories}
\author[a]{Stefano Cremonesi,}
\author[a]{Amihay Hanany,}
\author[b]{Noppadol Mekareeya,}
\author[c,d]{\\and Alberto Zaffaroni}

\affiliation[a]{Theoretical Physics Group, Imperial College London, \\
Prince Consort Road, London, SW7 2AZ, UK}
\affiliation[b]{Theory Division, Physics Department, CERN, \\CH-1211, Geneva 23, Switzerland}
\affiliation[c]{Dipartimento di Fisica, Universit\`a di Milano-Bicocca,  \\ I-20126 Milano, Italy}
\affiliation[d]{INFN, sezione di Milano-Bicocca, I-20126 Milano, Italy}

\emailAdd{s.cremonesi@imperial.ac.uk}
\emailAdd{a.hanany@imperial.ac.uk}
\emailAdd{noppadol.mekareeya@cern.ch}
\emailAdd{alberto.zaffaroni@mib.infn.it}

\preprint{
{\small
\begin{flushright}
CERN-PH-TH/2014-036\\
IMPERIAL-TP-14-SC-02
\end{flushright}
}
}

\abstract{
We evaluate the Coulomb branch Hilbert series of mirrors of  three dimensional Sicilian theories, which  arise from compactifying the $6d$ $(2,0)$ theory with symmetry $G$ on a circle times a Riemann surface with punctures. We 
obtain our result by gluing together the Hilbert series for building blocks $T_{\vec \rho}(G)$, where  $\vec \rho$ is a certain partition related to the dual group of $G$, which we evaluated in a previous paper. The result is expressed in terms of a class of symmetric functions, the Hall-Littlewood polynomials.  As expected from mirror symmetry, our results agree at genus zero with the superconformal index prediction for the Higgs branch Hilbert series of the Sicilian theories and extend it to higher genus. In the $A_1$ case at genus zero, we also evaluate the Coulomb branch Hilbert series of the Sicilian theory  itself, showing that it only depends on  the number of external legs. 

}

\begin{document}
\maketitle

\section{Introduction}

A general formula for  computing the generating function (Hilbert series) for the chiral ring associated with the Coulomb branch of three dimensional $\cN=4$ gauge theories  has been recently proposed \cite{Cremonesi:2013lqa}.  The formula counts monopole operators dressed by classical operators  and includes quantum corrections.  It can be applied to  any  $3d$ $\cN=4$ supersymmetric gauge theories that possess a Lagrangian description and that are good or ugly in the sense of \cite{Gaiotto:2008ak}. The formula has been successfully tested  against mirror symmetry in many cases \cite{Cremonesi:2013lqa,Cremonesi:2014kwa}.

In a companion paper we developed a machinery for  computing Coulomb branch Hilbert series  for wide classes of $\cN=4$ gauge theories by using  gluing techniques. We computed the Coulomb branch Hilbert series with background fluxes for the flavor symmetry of the three dimensional superconformal field theories  known as $T_{\vec \rho}(G)$ \cite{Gaiotto:2008ak}, a class of linear quiver theories with non-decreasing ranks associated with a partition $\vec \rho$ and a flavor symmetry $G$. We found an intriguing connection with a class of symmetric functions, the Hall-Littlewood polynomials, which have also appeared in the recent literature in the context of the superconformal index of four dimensional $\cN = 2$  theories \cite{Gadde:2011uv}.  We clarify the meaning of  this connection in the following. The $T_{\vec \rho}(G)$ theories  serve as  basic building blocks for constructing more complicated theories. 

 In this paper  we consider  the theories that arise from compactifying the $6d$ $(2,0)$ theory with symmetry $G=SU(N),SO(2N)$ on a circle times a Riemann surface with punctures. These are known as three dimensional Sicilian theories.  With the exception of the $SU(2)$ case, they have no Lagrangian description \cite{Gaiotto:2009we}. We are interested in their mirror which  can be obtained as follows. Starting from a set of  building blocks $\{T_{\vec \rho_1}(G), T_{\vec \rho_2}(G), \ldots, T_{\vec \rho_n}(G)\}$, one can construct  a new theory by gauging the common centerless flavor symmetry $G/Z(G)$, where $Z(G)$ is the center of $G$. 
We refer to this procedure as `gluing' the building blocks together.  The resulting theory is  the aforementioned mirror of the theory associated to a sphere with punctures $\{\vec \rho_1, \vec \rho_2, \ldots, \vec \rho_n \}$ \cite{Benini:2010uu, Nishioka:2011dq}.  

The main purpose of this paper is to compute the Coulomb branch Hilbert series of these mirrors. We do this by gluing together the Hilbert series of the theories $T_{\vec \rho_i}(G)$ as explained in \cite{Cremonesi:2014kwa}.
By mirror symmetry,  our results should agree with the Higgs branch Hilbert series of the Sicilian theories. The latter  can be computed for the four dimensional version of the theory,   since the Higgs branch of a theory with eight supercharges is protected against quantum corrections by a non-renormalization theorem \cite{Argyres:1996eh} and therefore is the same in all dimensions. Although the theory is non-Lagrangian, at genus zero the Higgs branch Hilbert series  can be written in terms of the Hall-Littlewood indices proposed in \cite{Gadde:2011uv, Gaiotto:2012uq}. We find perfect agreement  with the results in \cite{Gadde:2011uv, Gaiotto:2012uq,Lemos:2012ph}, as predicted by mirror symmetry.  

Our result clarifies why the Hall-Littlewood polynomials appear in two different contexts,  the Coulomb branch Hilbert series for the $T_{\vec \rho}(G)$ theories and a limit of the four dimensional superconformal index of Sicilian theories.   It is interesting to observe how the structure of the superconformal index formula (see for example \eref{genus0}), obtained in a completely different manner,  can be naturally reinterpreted in terms of gluing of three dimensional building blocks. 

Our gluing formula easily extends to punctured Riemann surfaces of higher genus, by incorporating adjoint hypermultiplets in the mirror theory \cite{Benini:2010uu}. For such Riemann surfaces the Hall-Littlewood index of the 4d non-Lagrangian theory differs from the Higgs branch Hilbert series, as discussed in \cite{Gadde:2011uv}. 
Our formula for the Coulomb branch Hilbert series of the 3d Lagrangian mirror theory \eqref{genSicilian} provides the Higgs branch Hilbert series of the (3d or 4d) non-Lagrangian Sicilian theory, for any genus $g$, as long as the theory is not \emph{bad} in the sense of \cite{Gaiotto:2008ak}. In section  \ref{sec:star2}, we successfully test our higher genus prediction for the case of $A_1$ Sicilian theories, also known as tri-vertex theories. 
 These are $3d$ $\CN=4$ Lagrangian theories associated to a graph with tri-valent vertices, where a finite line denotes an $SU(2)$ gauge group, an infinite line denotes an $SU(2)$ global symmetry, and a vertex denotes $8$ half-hypermultiplets in the tri-fundamental representation of $SU(2)^3$.  These graphs are characterized by the genus $g$ and the number of external legs $e$.  The Higgs branch Hilbert series of such theories were computed directly in \cite{Hanany:2010qu}. In section \ref{sec:generalgande} we reproduce that result from the Coulomb branch of the mirror theory.  

As an addition to the main line of this paper, which focusses on the Coulomb branch of mirrors of three dimensional Sicilian theories, in section \ref{sec:triv} we study the Coulomb branch of tri-vertex theories themselves at genus zero using the monopole formula.  We find that the Coulomb branch Hilbert series depends only on  $e$ and not on the details of the graph, as suggested in \cite{Benini:2010uu}.  

The paper is organized as follows. In section \ref{sec:general} we review  the formula for the Coulomb branch Hilbert series of ${\cal N}=4$  theories,  the gluing technique and the Hall-Littlewood formula
for the $T_{\vec \rho}(G)$ theories. In section \ref{sec:Sicilian} we compute the Coulomb branch Hilbert series of the mirrors of three dimensional Sicilian theories with A-type punctures at arbitrary genus $g$.
We examine in particular the case of the $T_N$ theory.  We successfully compare the result for genus zero with the  superconformal index prediction for Higgs branch Hilbert series of the Sicilian theories  
given in \cite{Gadde:2011uv, Gaiotto:2012uq}. We also give explicit examples for theories at higher genus. In the $SU(2)$ case, where the Sicilian theories are Lagrangian, we compare our result  with the Higgs branch Hilbert series  computed in \cite{Hanany:2010qu} finding perfect agreement. In section \ref{sec:SicilianD} we extend the analysis to theories of type D. As a general check of our predictions, we demonstrate
the equivalence between $D_3$ and $A_3$ punctures and we compute the Coulomb branch Hilbert series for a set of $D_4$ punctures where the Higgs branch Hilbert series can be explicitly evaluated,
finding perfect agreement. In section \ref{sec:triv}  we compute the Hilbert series of the tri-vertex theories at genus zero showing that they only depend on the number of external legs. In section \ref{sec:genfunctrivertex} we present generating functions and recursive formulae, which are powerful tools for computing the Hilbert series of tri-vertex theories. Finally, in appendix \ref{app:twisted} we consider theories of type D with twisted punctures.

\paragraph{Note added:} One might ask whether there is any relation between the Coulomb branch Hilbert series that we study and the $3d$ superconformal index \cite{Kim:2009wb, Imamura:2011su, Krattenthaler:2011da}. Indeed, a recent work \cite{Razamat:2014pta} appeared after the submission of this paper, showing that the superconformal index of a $3d$ $\CN=4$ theory reduces to the Hilbert series in a particular limit.

\section{Coulomb branch Hilbert series of a 3d $\cN=4$ gauge theory}\label{sec:general}
Our main aim is to study the Coulomb branch of three dimensional $\cN=4$ gauge theories.  Classically, this branch is parameterized by the vacuum expectation values of the triplet of scalars in the $\CN = 4$ vector multiplets and by the vacuum expectation value of the dual photons, at a generic point where the gauge group is spontaneously broken to its maximal torus. This yields a HyperK\"ahler space of quaternionic dimension equal to the rank of the gauge group. The Coulomb branch is, however, not protected against quantum corrections and the associated chiral ring has a complicated structure involving monopole operators in addition to the classical fields in the Lagrangian.

A suitable quantum description of the chiral ring on the Coulomb branch is to replace the above description by monopole operators.  The gauge invariant BPS objects on the branch are monopole operators dressed by a product of a certain scalar field in the vector multiplet.   The spectrum of such BPS objects can be studied in a systematic way by computing their partition function, known as the {\it Hilbert series}. A Hilbert series is a generating function of the chiral ring, which enumerates gauge invariant BPS operators which have a non-zero expectation value along the Coulomb branch. As extensively discussed in \cite{Cremonesi:2013lqa, Cremonesi:2014kwa}, a general formula for the Hilbert series of the Coulomb branch of an $\cN=4$ theory can be computed based on this principle.  We refer to such a formula as the {\it monopole formula}. 

In  \cite{Cremonesi:2014kwa} we found  an analytic  expression for  the Coulomb branch Hilbert series of a class of theories called $T_{\vec \rho}(G)$ \cite{Gaiotto:2008ak}, where $G$ is a classical group and $\vec \rho$ is a partition associated with the GNO dual group $G^\vee$.  Such a theory has a Lagrangian description \cite{Gaiotto:2008ak, Benini:2010uu, Cremonesi:2014kwa}.  The Hilbert series of these theories can be conveniently written in terms of Hall-Littlewood polynomials  \cite{Cremonesi:2014kwa}, and the corresponding formula is dubbed the {\it Hall-Littlewood formula}.  In the following section we show that the Hall-Littlewood formula is a convenient tool to compute the Coulomb branch Hilbert series of mirrors of three dimensional Sicilian theories.

Let us now summarize important information on the monopole and Hall-Littlewood formulae for Coulomb branch Hilbert series.

\subsection{The monopole formula} 
The monopole  formula  \cite{Cremonesi:2013lqa} counts all gauge invariant chiral operators that can acquire a non-zero expectation value along the Coulomb branch, according to their  dimension and quantum numbers. 
The operators are written in an  $\cN=2$ formulation and the $\cN=4$ vector multiplet is decomposed   into an $\cN=2$ vector multiplet and a chiral multiplet  $\Phi$ transforming in the adjoint representation of the gauge group. We refer to  \cite{Cremonesi:2013lqa} for an explanation of the formula and simply quote the final result here.

The  formula for a good or ugly  \cite{Gaiotto:2008ak} theory  with gauge group $G$  reads
\be\label{Hilbert_series}
H_G(t,z)=\sum_{\vec m\,\in\, \Gamma_{G^\vee}/W_{G^\vee}} z^{J(\vec m)} t^{\Delta(\vec m)} P_G(t;\vec m) \;.
\ee
The sum is over the magnetic charges of the monopoles $m$ which, up to a gauge transformation,  belong to  a Weyl Chamber of the weight lattice $\Gamma_{G^\vee}$ of the GNO dual group \cite{Goddard:1976qe}. $P_G(t;m)$ is a factor which counts the gauge invariants of the gauge group $H_{\vec m}$ unbroken by the monopole $m$ made with the adjoint scalar field $\phi$ in the multiplet $\Phi$, according to their dimension. It is given by 
\be\label{classical_dressing}
P_G(t; \vec m)=\prod_{i=1}^r \frac{1}{1-t^{d_i(\vec m)}} \;,
\ee
where $d_i(\vec m)$, $i=1,\dots,{\rm rank}\; H_{\vec m}$ are the degrees of the independent Casimir invariants of  $H_{\vec m}$.   $\Delta(m)$ is the quantum dimension of the monopole which is given by \cite{Borokhov:2002cg,Gaiotto:2008ak,Benna:2009xd,Bashkirov:2010kz}
\be\label{dimension_formula}
\Delta(\vec m)=-\sum_{\vec \alpha \in \Delta_+(G)} |\alpha(\vec m)| + \frac{1}{2}\sum_{i=1}^n\sum_{\vec \rho_i \in R_i}|\vec \rho_i(\vec m)|\;,
\ee
where $\vec \alpha$ are the positive roots of $G$  and $\vec \rho_i \in R_i$  the weights of the matter field representation $R_i$ under the gauge group.  $z$ is a fugacity valued in the topological symmetry group, which exists if $G$ is not simply connected,  and $J(\vec m)$ the topological charge of a monopole operator of GNO charges $\vec m$.

\paragraph{Turning on background magnetic fluxes.} 
As discussed in \cite{Cremonesi:2014kwa}, the formula can be generalized to include  background monopole fluxes for a global flavor symmetry $G_F$ acting on the matter fields: 
 \be\label{Hilbert_seriesBackground}
H_{G,G_F}(t,{\vec m_F},z)=\sum_{\vec m\,\in\, \Gamma_{G^\vee}/W_{G^\vee}}  z^{J(\vec m)} t^{\Delta(\vec m, \vec {\vec m_F})} P_G(t;\vec m) \;.
\ee
The sum  is only over the magnetic fluxes of the gauge group $G$ but depends on the  weights  ${\vec m_F}$ of the dual group $G^\vee_F$ which enter explicitly in the dimension formula (\ref{dimension_formula}) through all the matter fields that are charged under the global symmetry $G_F$. By using the global symmetry we can restrict the value of ${\vec m_F}$  to a Weyl chamber of $G^\vee_F$ and take ${\vec m_F}\in \Gamma_{G^\vee_F}/W_{G^\vee_F}$.

\paragraph{The gluing technique.} 
We can construct more  complicated theories by  starting  with a collection of theories and gauging  some common global symmetry $G_F$ they share. 
The Hilbert series of the final theory where $G_F$ is gauged  is given by multiplying the Hilbert series with background fluxes for $G_F$ of the building blocks, summing over the monopoles of $G_F$ and
including the contribution to the dimension formula of the $\cN=4$ dynamical vector multiplets associated with the gauged group $G_F$:
\be\label{gluing}
H(t)=\sum_{{\vec m_F}\,\in\, \Gamma_{G^\vee_F}/W_{G^\vee_F}} t^{-\sum_{\vec \alpha_F \in \Delta_+(G_F)} |\alpha_F({\vec m_F})|} P_{G_F}(t;{\vec m_F}) \prod_i  H^{(i)}_{G,G_F}(t,{\vec m_F})\; ,
\ee
where $\alpha_F$ are the positive roots of $G_F$ and the product with the index $i$ runs over the Hilbert series of the $i$-th theory that is taken into the gluing procedure. 
Since we can always make $\alpha_F({\vec m_F})$ non-negative by choosing ${\vec m_F}$ in the main Weyl chamber, the evaluation of $H(t)$ turns out to have no absolute values.
The formula  (\ref{gluing}) can be immediately generalized to include fugacities for the topological symmetries acting on the Coulomb branch.  

In the next sections we will provide explicit and general formulae for many interesting 3d ${\cal N}=4$  superconformal theories including mirrors of M5-brane theories compactified on a circle times a Riemann surfaces with punctures. They are obtained by gluing a simple class of building blocks that we now discuss.

\subsection{The Hall-Littlewood formula}
As extensively discussed in \cite{Cremonesi:2014kwa}, the Coulomb branch Hilbert series of $T_{\vec \rho}(G^\vee)$ for a classical group $G$ can be computed using formulae involving Hall-Littlewood polynomials.  The main purpose of this paper is to show that these formulae are useful for computing Coulomb branch Hilbert series of mirrors of $3d$ Sicilian theories.  For the sake of completeness of the paper, we review Hall-Littlewood formulae below.   We first present the formula for $G=SU(N)$ and then discuss the formula for other classical groups, namely $SO(N)$ and $USp(2N)$.

\subsubsection{$T_{\vec \rho} (SU(N))$}
The quiver diagram for $T_{\vec \rho} (SU(N))$ is
\bea
[U(N)]-(U(N_1))-(U(N_2))- \cdots -(U(N_d)), \label{quiv:TrhoG}
\eea
where the partition $\vec \rho$ of $N$ is given by
\bea \label{rhopartition}
\vec \rho = (N-N_1, N_{1}-N_{2}, N_{2} - N_{3}, \ldots, N_{d-1}-N_d, N_d)~,
\eea
with the restriction that $\vec \rho$ is a non-increasing sequence:
\bea
N-N_1 \geq N_{1}-N_{2} \geq N_{2} - N_{3} \geq \cdots \geq N_{d-1}-N_d \geq N_d > 0~. \label{resTrhoSUN}
\eea
The quiver theory in \eref{quiv:TrhoG} can be realised from brane configurations as proposed in \cite{Hanany:1996ie}.

The Coulomb branch Hilbert series of this theory can be written as
\be \label{HLformulaT}
\begin{split}
&H[{T_\rho (SU(N))}] (t; x_1, \ldots, x_{d+1} ; n_1, \ldots, n_N) \\
&= t^{\frac{1}{2} \delta_{U(N)}(\vec n)}  (1-t)^N K^{U(N)}_{\vec \rho} (\vec x;t)\Psi_{U(N)}^{\vec n}(\vec x t^{\frac{1}{2}\vec w_{\vec \rho}}; t) ~,
\end{split}
\ee
where the Hall-Littlewood polynomial associated with the group $U(N)$ is given by
\bea
\Psi^{\vec n}_{U(N)} (x_1,\dots,x_N;t)=\sum_{\sigma \in S_N}
x_{\sigma(1)}^{n_1} \dots x_{\sigma(N)}^{n_N}
\prod_{1 \leq i<j \leq N}   \frac{  1-t x_{\sigma(i)}^{-1} x_{\sigma(j)} } {1-x_{\sigma(i)}^{-1} x_{\sigma(j)}}~,
\eea
with $n_1, \ldots, n_N$ the background GNO charges for $U(N)$ group, with 
\bea n_1 \geq n_2 \geq \cdots \geq n_N \geq 0~.\eea
The notation $\delta_{U(N)}$ denotes the sum over positive roots of the group $U(N)$ acting on the {\it background} charges $n_i$:
\bea
\delta_{U(N)}(\vec n) = \sum_{1\leq i < j \leq N} (n_i - n_j) = \sum_{j=1}^{N} (N+1-2j) n_j~.
\eea
The fugacities $x_1, \ldots, x_{d+1}$ are subject to the following constraint which fixes the overall $U(1)$:
\bea \label{constrx}
\prod_{i=1}^{d+1} x_i^{\rho_i} =1~.
\eea
The vector $\vec w_{{r}}$ denotes the weights of the $SU(2)$ representation of dimension $r$:
\bea
\vec w_{{r}} = (r-1, r-3, \ldots, 3-r, 1-r)~.
\eea
Hence the notation $t^{\frac{1}{2}\vec w_{r}}$ represents the vector
\bea
t^{\frac{1}{2}\vec w_{r}} = (t^{\frac{1}{2}(r-1)}, t^{\frac{1}{2}(r-3)}, \ldots, t^{-\frac{1}{2}(r-3)},t^{-\frac{1}{2}(r-1)})~.
\eea
In \eref{HLformulaT} and henceforth, we abbreviate
\bea
\Psi_{U(N)}^{\vec n}(\vec x t^{\frac{1}{2}\vec w_{\vec \rho}}; t) := \Psi_{U(N)}^{(n_1, \ldots, n_N)}(x_1 t^{\frac{1}{2}\vec w_{\rho_1}}, x_2 t^{\frac{1}{2}\vec w_{\rho_2}} , \ldots, x_{d+1} t^{\frac{1}{2}\vec w_{\rho_{d+1}}};t)~.
\eea
The prefactor $K^{U(N)}_{\vec \rho} (\vec x;t)$  is given by
\bea \label{KUN}
K^{U(N)}_{\vec \rho} (\vec x;t) = \prod_{i=1}^{\text{length}({\vec \rho}^T)} \prod_{j,k=1}^{\rho^T_i} \frac{1}{1-a^i_j \bar{a}^i_k}~,
\eea
where $\rho^T$ denotes the transpose of the partition $\rho$ and we associate the factors
\be
\begin{split}
a^i_j &= x_j \;\; t^{\frac{1}{2} (\rho_j-i+1)}~, \qquad  i=1,\dots,\rho_j \\
{\bar a}^i_k &= x_k^{-1} t^{\frac{1}{2} (\rho_k-i+1)}~, \qquad  i=1,\dots,\rho_k \label{defaabar}
\end{split}
\ee
to each box in the Young tableau. The powers of $t$ inside $a^i_j$ and ${\bar a}^i_k$ are positive by construction.

We demonstrate the HL formula \eref{HLformulaT} in a number of examples in section \ref{sec:Sicilian}.

\subsubsection{$T_{\vec \rho} (G^\vee)$}
In this  section we review a generalized version of formula \eref{HLformulaT} to a more general classical group $G$.  The quiver diagrams are explicitly given in \cite{Cremonesi:2014kwa}. Further discussions regarding mathematical aspects of this formula can be found in \cite{Mekareeya:2012tn, Chacaltana:2012zy, Cremonesi:2014kwa}.

The partition $\vec \rho$ induces an embedding $\vec \rho: \mathrm{Lie} (SU(2)) \rightarrow  \mathrm{Lie} (G)$ such that
\bea
[1,0,\ldots,0]_G =  \bigoplus_i [\rho_i-1]_{SU(2)}~.
\eea
The global symmetry $G_{\vec \rho}$ associated to the puncture $\vec \rho = [\rho_i]$, with $r_k$ the number of times that part $k$ appears in the partition $\vec \rho$, is given by
\bea \label{sympunc}
G_{\vec \rho} = \begin{cases} S \left( \prod_{k} U(r_k)  \right) & \qquad G= U(N)~, \\
 \prod_{k~\text{odd}} SO(r_k) \times \prod_{k~\text{even}} USp(r_k) & \qquad G= SO(2N+1)~\text{or}~SO(2N)~, \\
 \prod_{k~\text{odd}} USp(r_k) \times \prod_{k~\text{even}} SO(r_k) & \qquad G= USp(2N)~.
\end{cases}
\eea

Let $x_1, x_2, \ldots$ be fugacities for the global symmetry $G_{\vec \rho}$, the commutant of $\rho(SU(2))$ in $G$, and $r(G)$ the rank of $G$.  In \cite{Cremonesi:2014kwa} we have conjectured that the Coulomb branch Hilbert series is given by the HL formula
\be
H[T_{\vec \rho}(G^\vee)](t; \vec x;n_1,\ldots, n_{r(G)}) = t^{\frac{1}{2} \delta_{G^\vee}(\vec n)} (1-t)^{r(G)} K^{G}_{\vec \rho}  (\vec x;t) \Psi^{\vec n}_{G}(\vec a(t, \vec x);t)~. \label{mainHL}
\ee
Here $\Psi^{\vec n}_{G}$ is the Hall-Littlewood polynomial associated to a Lie group $G$, given by
\bea
\Psi^{\vec n}_{G} (x_1, \ldots, x_r; t) = \sum_{w \in W_G} {\vec x}^{w(\vec n)}
\prod_{\vec \alpha \in \Delta_+(G)}  \frac{  1-t {\vec x}^{-w({\vec \alpha})} } {1- {\vec x}^{-w({\vec \alpha})} }~,
\eea
where $W_G$ denotes the Weyl group of $G$, $\Delta_+(G)$ the set of positive roots of $G$,  $\vec n= \sum_{i=1}^r n_i \vec{e}_i$, with $\{ \vec e_1, \ldots, \vec e_r\}$ the standard basis of the weight lattice and $r$ the rank of $G$. See Appendix B of \cite{Cremonesi:2014kwa} for more details.  $G^\vee$ is the GNO  dual group \cite{Goddard:1976qe}. The power $\delta_{G^\vee}(\vec n)$ is the sum over positive roots $\vec \alpha\in\Delta_+(G^\vee)$  of the flavor group $G^\vee$ acting on the {\it background} monopole charges $\vec n$: 
\bea
\delta_{G^\vee}(\vec n) = \sum_{\vec \alpha \in \Delta_+(G^\vee)} |\vec\alpha(\vec n)|~.
\eea
Explicitly, for classical groups $G$ and fluxes $\vec{n}$ in the fundamental Weyl chamber, these are given by
\bea \label{powers}
\delta_{G^\vee} (\vec n) &= \begin{cases} 
\sum_{j=1}^N (N+1-2j) n_j & \qquad G^\vee =G=U(N), \\
\sum_{j=1}^{N} (2N+1-2j)n_j & \qquad G^\vee=B_N,~ G=C_N\\
 \sum_{j=1}^{N} (2N+2-2j)n_j   & \qquad G^\vee=C_N,~ G=B_N \\
\sum_{j=1}^{N-1} (2N-2j)n_j  & \qquad G^\vee =G=D_N~.
\end{cases}
\eea

The argument $\vec a(t, \vec x)$ of the HL polynomial, which we shall henceforth abbreviate as $\vec a$, is determined by the following decomposition of the fundamental representation of $G$ to $G_{\vec \rho} \times {\vec \rho} (SU(2))$:
\bea \label{decompfund}
\chi^G_{{\bf fund}} (\vec a) = \sum_{k}  \chi^{G_{\rho_k}}_{{\bf fund}}( \vec x_k) \chi^{SU(2)}_{[\rho_k -1]} (t^{1/2})~,
\eea
where $G_{\rho_k}$ denotes a subgroup of $G_{\vec \rho}$ corresponding to the part $k$ of the partition $\vec \rho$ that appears $r_k$ times.  Formula \eref{decompfund} determines $\vec a$ as a function of $t$ and $\{ \vec x_k \}$ as required.  Of course, there are many possible choices for $\vec a$; the choices that are related to each other by outer-automorphisms of $G$ are equivalent.

The prefactor $K^G_{\vec \rho}(\vec x; t)$ is independent of ${\vec n}$ and can be determined as follows.  The embedding specified by $\vec \rho$ induces the decomposition 
\begin{equation}
\chi^G_{\bf Adj} (\vec a) = \sum_{j =0, \frac{1}{2}, 1, \frac{3}{2}, \ldots}  \chi^{G_{\vec \rho}}_{R_j}(\vec x_j)  \chi^{SU(2)}_{[2j]}(t^{1/2})~, \label{decompadj} 
\end{equation} 
where $\vec a$ on the left hand side is the same $\vec a$ as in \eref{decompfund}.  Each term in the previous formula gives rise to a plethystic exponential,%
\footnote{See Appendix A of \cite{Cremonesi:2013lqa} for the definition.} giving 
\begin{equation}
K^G_{\vec \rho}(\vec x; t)=\PE \left[\sum_{j =0, \frac{1}{2}, 1, \frac{3}{2}, \ldots}  t^{j+1} \chi^{G_{\vec \rho}}_{R_j}({\vec x}_j )\right].  \label{K}
\end{equation}

As a remark, in the special case when $\vec \rho: \mathrm{Lie} (SU(2)) \rightarrow  \mathrm{Lie} (G)$ is a \emph{principal embedding} $\vec\rho_{princ}$ (see, \eg~\cite{Chacaltana:2012zy}), the global symmetry acting on the Coulomb branch is trivial $G_{\vec\rho_{princ}}=1$, the prefactor  $K^G_{\vec \rho_{princ}}(t)=P_{G}(t;\vec 0)$ equals the Casimir factor of $G$ (or equivalently $G^\vee$), and the Hall-Littlewood formula \eref{mainHL} reduces to
\be\label{HL_princ}
H[T_{\vec \rho_{princ}}(G^\vee)](t; n_1,\ldots, n_{r(G)}) = 1~. 
\ee
This identity has a simple physical interpretation in the context of mirrors of Sicilian theories that we consider in this paper: adding an \emph{empty puncture} does not affect the Hilbert series of the Sicilian theory.

In the following we discuss several examples of mirrors of $3d$ Sicilian theories for which we use the Hall-Littlewood formulae to compute their Coulomb branch Hilbert series.

\section{Mirrors of 3d Sicilian theories of  $A$-type} \label{sec:Sicilian}
In this and the next section  we evaluate the Coulomb branch Hilbert series of the mirror of the theories arising from compactifying the $6d$ $(2,0)$ theory with symmetry $G$ on a circle times a Riemann surface with punctures,  also called  Sicilian theories. These theories and their Coulomb branch Hilbert series will be obtained  by gluing together $T_{\vec \rho}(G)$ building blocks. 

\begin{figure}[H]
\begin{center}
\includegraphics[scale=0.5]{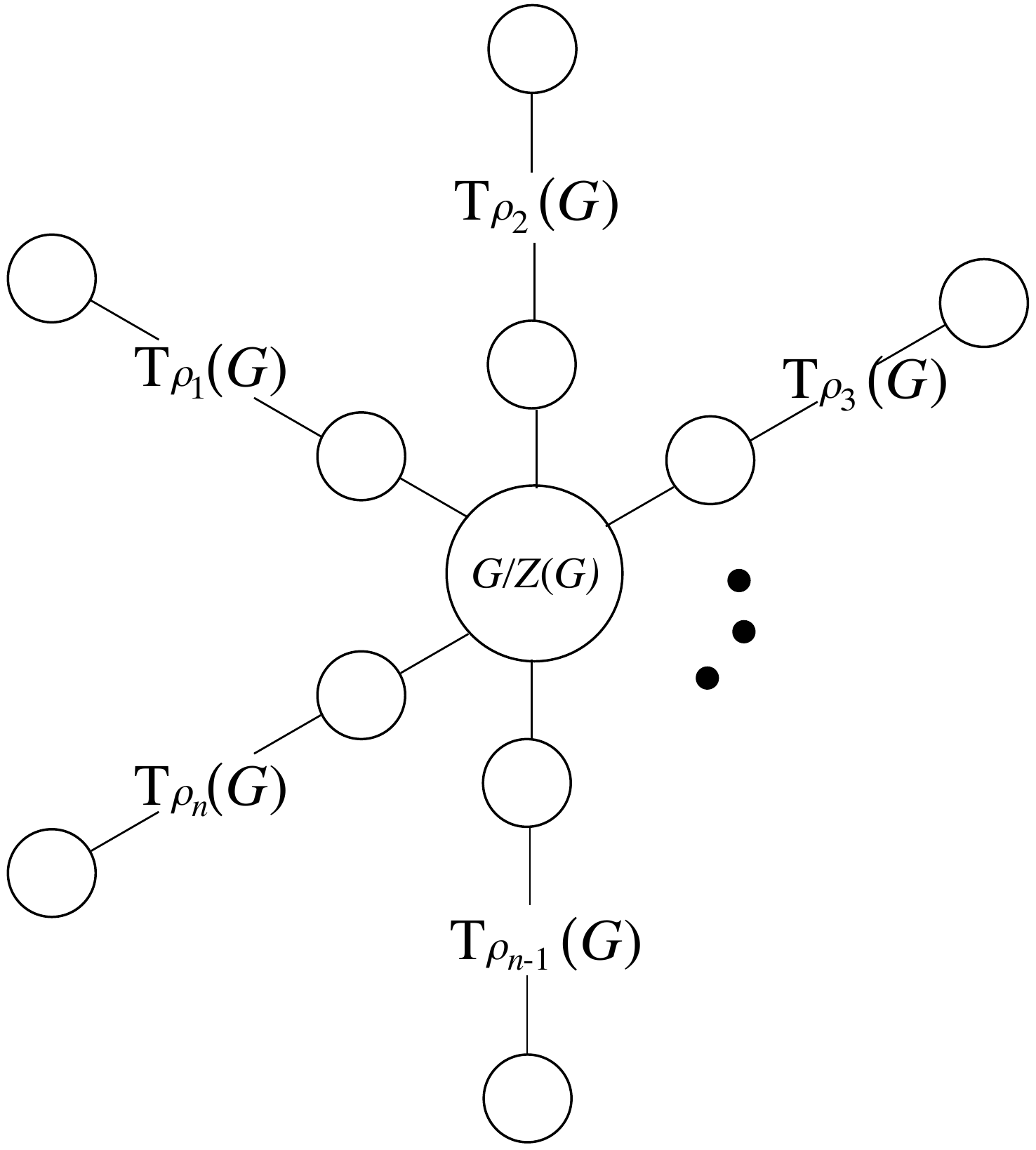}
\caption{Gluing $T_{\vec \rho_1}(G), \ldots, T_{\vec \rho_n}(G)$ via the common centerless flavor symmetry $G/Z(G)$.  This is a mirror theory of the theory on M5-brane compactified on a circle times a Riemann sphere with punctures $\vec \rho_1, \ldots, \vec \rho_n$.}
\label{fig:gluing}
\end{center}
\end{figure}

Given a set of theories $\{T_{\vec \rho_1}(G), \ldots, T_{\vec \rho_n}(G)\}$, we can construct a new theory by gauging the common centerless flavor symmetry $G/Z(G)$; see \fref{fig:gluing}.%
\footnote{For $G=U(N)$, this also involves factoring out a decoupled $U(1)$ gauge group.} The resulting theory is the mirror  of the theory on M5-branes wrapping a circle times a Riemann sphere with punctures $\vec \rho_1, \ldots, \vec \rho_n$ \cite{Benini:2010uu, Nishioka:2011dq}. 
For example, taking $G=SU(3)$ and $\vec \rho_1=\vec \rho_2=\vec \rho_3 =(1,1,1)$ we obtain a mirror of the $T_3$ theory reduced to three dimensions.  
Recall that the Higgs branch of the 3d $T_3$ theory is the reduced moduli space of $1$ $E_6$ instantons on $\BC^2$ and the Coulomb branch is $\BC^2/{\hat{E}_6}$. 
The moduli spaces of $k$ $E_6$, $E_7$ and $E_8$ instantons on $\BC^2$ can be also realized as the Higgs branch of the $6d$ $(2,0)$ theory compactified on a circle times a Riemann sphere with punctures.

We demonstrate how to `glue' the Hilbert series  $T_{\vec \rho_n}(G)$ together to obtain the Coulomb branch Hilbert series of the mirror of the theory on M5-branes compactified on $S^1$ times a Riemann sphere with punctures $\{ \vec \rho_i \}$.  By mirror symmetry, this is equal to the Higgs branch Hilbert series of the latter. The theories on M5-branes are not Lagrangian, but, when the genus of the Riemann surface is zero, the Higgs branch Hilbert series  can be evaluated  by the Hall-Littlewood (HL) limit of the superconformal index \cite{Gadde:2011uv}.  We find perfect agreement with the results in \cite{Gadde:2011uv}, which were obtained in a completely different manner. Upon introduction of $g$ $G$-adjoint hypermultiplets \cite{Benini:2010uu}, our formulae can be used also for genus greater than one, where the  Higgs branch Hilbert series for the M5-brane theories cannot be evaluated as a limit of the $4d$ superconformal index.%
\footnote{For genus greater than $1$, the F-terms of the theory are not all independent. As a result, the HL limit of the $4d$ superconformal index fails to reproduce the Higgs branch Hilbert series. For a very clear explanation of this technical fact, see section 5 of \cite{Gadde:2011uv}.}
 In section \ref{sec:twoM5} we  will be able to test the validity of our result for higher genus in the case of two M5 branes where the theory is Lagrangian and we can use conventional methods for computing the Higgs branch Hilbert series. 

In this section we discuss the case of A-type theories with $G=SU(N)$ and in the next section
we discuss D-type theories with $G=SO(2N)$.

\subsection{Mirrors of tri-vertex theories: star-shaped  $U(2)\times U(1)^e/U(1)$ quivers}\label{sec:twoM5}\label{sec:star2}

\begin{figure}[htbp]
\begin{center}
\includegraphics[scale=0.4]{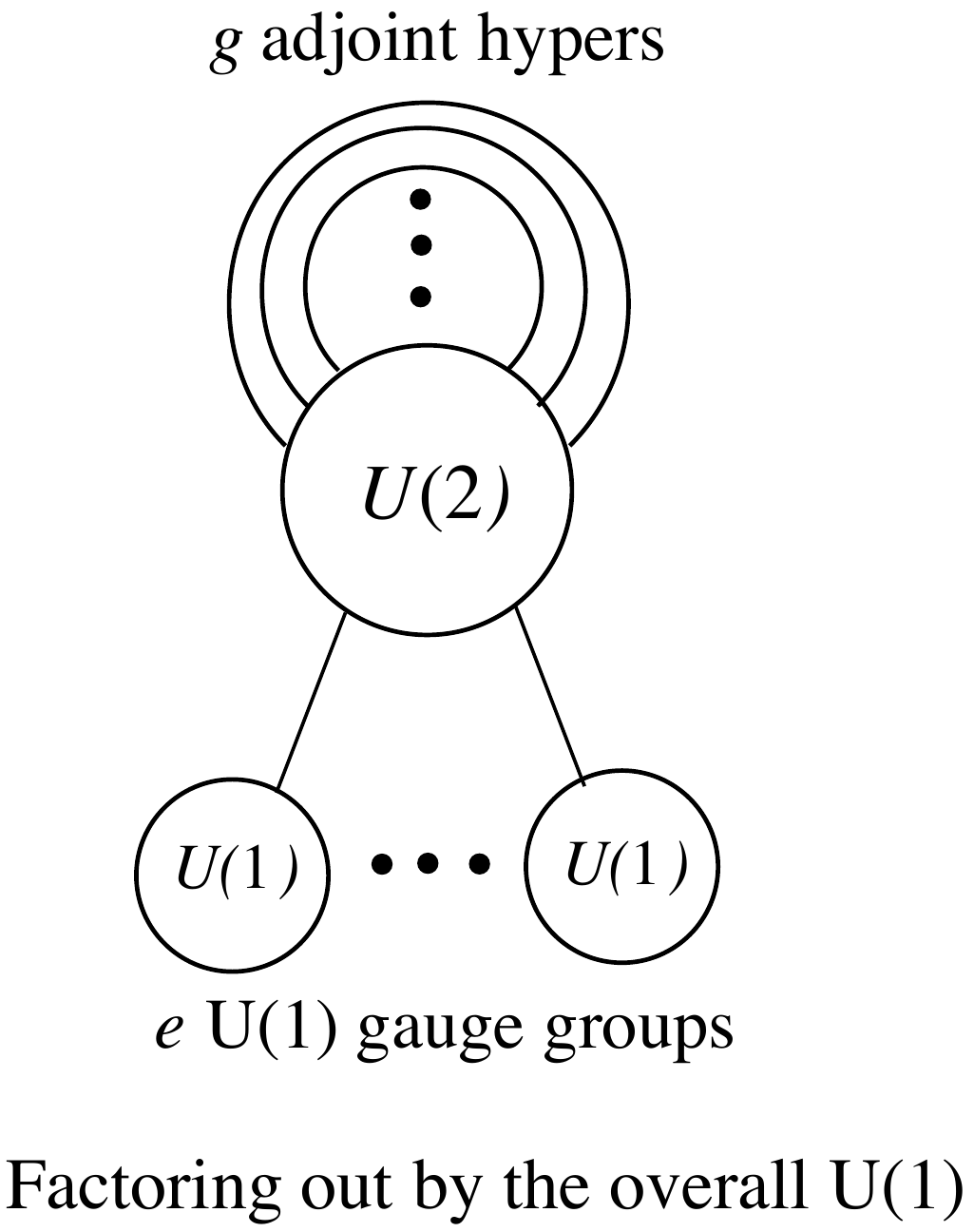}
\caption{The mirror theory of a tri-vertex theory with genus $g$ and $e$ external legs.}
\label{fig:MirrorTrivert}
\end{center}
\end{figure}

We start by considering the Coulomb branch Hilbert series of the mirrors of  theories on {\it two} M5-branes compactified on a circle times a Riemann surface with punctures. The latter are referred to as $3d$ $SU(2)$ Sicilian theories \cite{Benini:2009mz, Benini:2010uu} or $3d$ theories with tri-vertices \cite{Hanany:2010qu}. They are Lagrangian theories  whose quiver is explicitly discussed in section \ref{sec:triv}. According to \cite{Benini:2010uu}, the  mirror of a tri-vertex theory with genus $g$ and $e$ external legs is a star-shaped  $U(2)\times U(1)^e/U(1)$ quiver gauge theory with the $U(2)$ node with $g$ adjoint hypermultiplets in the center, attached to $e\geq 3$ $U(1)$ nodes around it. The quiver is depicted in \fref{fig:MirrorTrivert}.

The overall $U(1)$ gauge group in the quiver is decoupled and needs to be factored out. It is crucial to mod out by the overall $U(1)$ properly: in particular the quiver with an $SU(2)$ node in the center, attached to $e$ $U(1)$ nodes around it, gives the wrong Coulomb branch, which disagrees with the Higgs branch of the $g=0$ tri-vertex theory with $e\geq 3$ legs \cite{Hanany:2010qu}. The reason is that $U(2)=U(1)\times SU(2)/\bZ_2$.

Let us first consider the $U(2)\times U(1)^e$ quiver gauge theory which includes the decoupled overall $U(1)$. We use GNO charges $n_1$ and $n_2$ for $U(2)$, related to the integer weights $n_1\geq n_2 > -\infty$ in the Weyl chamber. For the $i$-th $U(1)$ gauge group, with $i=1,\ldots, e$, we use the GNO charge
$m_i\in \bZ$. The dimension formula (\ref{dimension_formula}) reads
\bea\label{mondim_star2}
\Delta_g(n_1,n_2; m_1,\dots,m_e) 
= (g-1)|n_1-n_2| +\frac{1}{2} \sum_{i=1}^e ( |n_1-m_i| +|n_2-m_i| )~.
\eea
The formula is invariant under the common shift $n_{1,2} \to n_{1,2}+c$, $m_i \to m_i + c$, with $i=1,\dots,e$ and $c\in \bZ$: this is the decoupled $U(1)$ that we have to fix. 

\paragraph{Topological factor.}
The topological $U(1)_J^{e+1}$ fugacities for the naive $U(2)\times U(1)^e$ theory contribute 
\bea 
z_0^{n_1+n_2}\prod_{i=1}^e z_i^{m_i}~, \label{topologicale}
\eea
where $z_0$ is the fugacity associated to the topological charge of $U(2)$ and $z_i$, with $i=1,\ldots, e$, the fugacity associated to the topological charge of the $i$-th copy of $U(1)$.

\paragraph{Factoring out the overall $U(1)$.}
To get rid of the decoupled $U(1)$, which would make this a bad theory, we fix the $\bZ$ shift symmetry of the magnetic fluxes and impose a relation on the topological fugacities $z_I$, where $I=0,1,\dots,e$. Different fixings make manifest different topological symmetry enhancements. Here we want to manifest an $SU(2)^e$ enhanced topological symmetry, with one $SU(2)$ per external $U(1)$ node. Therefore we fix the overall $U(1)$ by imposing 
\be\label{fixingU(1)center}
n_2=0\;,\qquad\qquad z_0^2 \prod_{i=1}^e z_i=1\;.  
\ee
In the following we choose to write
\bea
z_0=\epsilon~ x_1\cdots x_e~, \qquad  z_i = x_i^{-2}~, \quad i =1, \ldots, e~, \qquad \epsilon^2=1 \label{mapzx}
\eea
As we shall see, this choice makes $SU(2)$ characters manifest in the Hilbert series.  $\epsilon$ is the fugacity of a potential discrete $\BZ_2$ topological symmetry. 
This $\BZ_2$ can be absorbed into the center of an $SU(2)$ symmetry, and correspondingly $\epsilon$ can be absorbed into $z_i$ or $x_i$, except for the case of no punctures $e=0$, where it is the topological symmetry for the gauge group $SU(2)/\BZ_2$. We will sometimes omit $\epsilon$ in the following.

\subsubsection*{The monopole formula for Coulomb branch Hilbert series}
Following the above discussion, the refined Hilbert series of the Coulomb branch (\ref{Hilbert_series}) reads
\bea\label{HS_Coulomb_star2}
& H[\text{mirror $(g, e)$}](t;x_1,\dots,x_e)  \\
&= \sum_{n_1 \geq n_2 =0}^\infty \sum_{m_i \in \bZ}  t^{\Delta_g(n_1,n_2; m_1,\dots,m_e)} P_{U(1)} (t)^e (1-t)P_{U(2)}(t;n_1,n_2) \epsilon^{n_1+n_2}\prod_{i=1}^e x_i^{n_1+n_2-2m_i} \;, \nn
\eea
where the last factor  comes from \eref{topologicale} and \eref{mapzx}.
The classical factors are given by 
\be P_{U(1)} (t)= \frac{1}{1-t}\ee
and 
\be\label{classical_U(2)}
P_{U(2)}(t;n_1,n_2)= \begin{cases}
\frac{1}{(1-t)(1-t^2)}\;, &\quad n_1=n_2 \\
\frac{1}{(1-t)^2}\;, &\quad n_1\ne n_2 
\end{cases}\;.
\ee
The factor $(1-t)$ in front of $P_{U(2)}$ removes the classical invariants of the decoupled $U(1)$. 

As we show explicitly in subsection \ref{sec:generalgande}, evaluating the monopole formula (\ref{HS_Coulomb_star2}) reproduces the refined Hilbert series of the Higgs branch of the mirror theory, formula (7.1) of \cite{Hanany:2010qu}, under the fugacity map $t_{here}=t^2_{there}$.

\subsubsection*{Coulomb branch Hilbert series from gluing}

It is instructive to rewrite \eref{HS_Coulomb_star2} as
\be \label{glueeT2}
\begin{split}
H[\text{mirror $(g, e)$}](t;x_1,\dots,x_e) &=   \sum_{n_1 \geq n_2 =0} (1-t) P_{U(2)} (t; n_1,n_2) t^{(g-1)(n_1-n_2)} \times  \\
& \quad \epsilon^{n_1+n_2} \prod_{j=1}^e H[T(SU(2))](t; x_j, x_j^{-1}; n_1,n_2)  ~,
\end{split}
\ee
where $H[T(SU(2))]$ is the Coulomb branch Hilbert series with background fluxes of the $T(SU(2))$ theory given by \eref{HLformulaT} with $\vec\rho =(1,1)$:
\be
\begin{split}
H[T(SU(2))]&(t; x, x^{-1}; n_1,n_2) = \sum_{m \in \BZ} t^{\frac{1}{2}(|m-n_1|+|m-n_2|)} x^{-2m} P_{U(1)} (t)  \\
&= t^{\frac{1}{2}(n_1-n_2)} (1-t)^2 \PE[ (1+[2]_x)t ] \Psi_{U(2)}^{(n_1,n_2)}(x,x^{-1};t)~.
\end{split}
\ee
Eq. \eref{glueeT2} is nothing but the gluing formula for the Coulomb branch Hilbert series of the star-shaped quiver, which results from gauging the common flavor symmetry of $e$ copies of $T(SU(2))$ and introducing $g$ adjoint hypermultiplets under the $U(2)$ group. The gluing factor is
\bea
(1-t) P_{U(2)} (t; n_1,n_2) t^{(g-1)|n_1-n_2|} \epsilon^{n_1+n_2} \prod_{j=1}^e  x_j^{n_1+n_2}~,
\eea
with $x_j^{n_1+n_2}$ factors already incorporated in $H[T(SU(2))]$ for convenience.

\subsubsection{Computation of the Hilbert series for general $g$ and $e$} \label{sec:generalgande}

We now compute the Coulomb branch Hilbert series \eqref{glueeT2} of the mirror of tri-vertex theories with genus $g$ and $e$ external legs. 

Using \eref{HLformulaT}, we obtain
\be \label{CoulombT11}
\begin{split}
H[T(SU(2))](t;  x,x^{-1} ; n, 0) &= t^{\frac{1}{2}n}  (1-t)^2 \PE[ (1+\chi^{SU(2)}_{[2]}(x) )t ] \Psi_{U(2)}^{(n,0)}(x,x^{-1};t) \\
&= t^{\frac{1}{2}n} (1-t) \PE[ t \chi^{SU(2)}_{[2]}(x) ] \Psi_{U(2)}^{(n,0)}(x,x^{-1};t)~,
\end{split}
\ee
where $[2]$ represents the adjoint representation of $SU(2)$.
An explicit formula for $\Psi_{U(2)}^{(n,0)}(x,x^{-1};t)$  is known in terms of $SU(2)$ characters:
\bea
\Psi_{U(2)}^{(n,0)}(x,x^{-1};t) = \chi^{SU(2)}_{[n]}(x) - t\chi^{SU(2)}_{[n-2]}(x)~,
\eea
where
\bea
\chi^{SU(2)}_{[n]}(x)  = \frac{x^{n+1}-x^{-(n+1)}}{x-x^{-1}}~,
\eea
which we extend to $n\in \bZ$.
Observe that $(1-t)\PE[ \chi^{SU(2)}_{[2]}(x)  t]\Psi_{U(2)}^{(m,0)}(x,x^{-1};t)$ is equal to the function $f_m(t,x)$ defined in (7.18) of \cite{Hanany:2010qu}:
\be
\begin{split}
f_m(t,x) &:= (1-t)\PE[ \chi^{SU(2)}_{[2]}(x)  t] \Psi_{U(2)}^{(m,0)}(x,x^{-1};t)  \\
&= (1-t) (\chi^{SU(2)}_{[m]}(x) -\chi^{SU(2)}_{[m-2]}(x) t) \PE[ [2]_x t] \\
&= \sum_{n=0}^\infty \chi^{SU(2)}_{[2n+m]}(x) t^{n}~.
\end{split}
\ee
Hence from \eref{CoulombT11} we have
\bea
H[T(SU(2))] (t;  x,x^{-1}; m, 0) &= t^{\frac{1}{2}m}  f_m(t,x)~.
\eea
Substituting this into \eref{glueeT2}, we obtain
\be
\begin{split}
& H[\text{mirror $(g, e)$}]  (t; x_1, \ldots, x_e; \epsilon)  \\
&=  \sum_{m=0}^\infty t^{\frac{1}{2}\chi m} \epsilon^m (1-t) P_{U(2)} (t;m,0) \prod_{j=1}^e f_m(t,x_j) \\
&=  \frac{1}{1-t^2}  \prod_{j=1}^e f_0(t,x_j) +\sum_{m=1}^\infty  \frac{t^{\frac{1}{2}\chi m} \epsilon^m}{1-t} \prod_{j=1}^e f_m(t,x_j)   \\
&= \frac{1}{1-t^2} \sum_{m=0}^\infty \left[ t^{\frac{1}{2} \chi m} \epsilon^m \prod_{j=1}^e f_m(t,x_j) + t^{\frac{1}{2}(\chi(m+1)+2)} \epsilon^{m+1} \prod_{j=1}^e f_{m+1}(t,x_j) \right]~,
\end{split}
\ee
where $\chi=2g+e-2$. This result precisely equals the Higgs branch Hilbert series of the mirror tri-vertex theory, (7.19) of \cite{Hanany:2010qu}, after the redefinition $t \rightarrow t^2$ and setting $\epsilon=1$.
Note that when $e>0$, the $\BZ_2$ topological symmetry can be absorbed into the center of any of the global $SU(2)$ factors, therefore we can set $\epsilon=1$. When $e=0$, $\epsilon$ is the fugacity for the actual $\BZ_2$ topological symmetry of the $SU(2)/\BZ_2$ theory with $g$ adjoint hypermultiplets. The Hilbert series of the Coulomb branch is 
\be
H[\text{mirror $(g,0)$}]  (t;\epsilon)= \PE[t^2 +\epsilon (t^{g-1}+t^g)-t^{2g}]~,
\ee
indicating a $\bC^2/\hat{D}_{g+1}$ singularity. The monopole generators of dimension $g-1$ and $g$ are odd under $\bZ_2$. This $\BZ_2$ symmetry acts on the Higgs branch of the mirror side by flipping sign to any one of the tri-fundamentals in the generators at page 27 of \cite{Hanany:2010qu}.

\subsection{The Coulomb branch of the mirror of $T_N$}

\begin{figure}[H]
\begin{center}
\begin{tikzpicture}[font=\scriptsize \bfseries]
\begin{scope}[auto,%
  every node/.style={draw, minimum size=1.2cm}, node distance=0.6cm];
\node[circle] (UN) at (0,0) {$N$};
\node[circle, right=of UN] (UNm1r) {$N-1$};
\node[circle, left=of UN] (UNm1l) {$N-1$};
\node[circle, above=of UN] (UNm1a) {$N-1$};
\node[circle, above=of UNm1a] (UNm2a) {$N-2$};
\node[circle, right=of UNm1r] (UNm2r) {$N-2$};
\node[circle, left=of UNm1l] (UNm2l) {$N-2$};
\node[draw=none, right=of UNm2r] (dotsr) {${\Large \mathbf \cdots}$};
\node[draw=none, left=of UNm2l] (dotsl) {${\Large \mathbf \cdots}$};
\node[draw=none, above=of UNm2a] (dotsa) {${\vdots}$};
\node[circle, right=of dotsr] (U1r) {$1$};
\node[circle, left=of dotsl] (U1l) {$1$};
\node[circle, above=of dotsa] (U1a) {$1$};
\end{scope}
\draw (UN)--(UNm1r)
(UN)--(UNm1l)
(UN)--(UNm1a)
(UNm1r)--(UNm2r)
(UNm2r)--(dotsr)
(UNm1l)--(UNm2l)
(UNm2l)--(dotsl)
(UNm1a)--(UNm2a)
(UNm2a)--(dotsa)
(dotsr)--(U1r)
(dotsl)--(U1l)
(dotsa)--(U1a);
\end{tikzpicture}
\end{center}
\caption{Quiver diagram for the mirror of $T_N$. Each node represents a unitary group of the labelled rank and the overall $U(1)$ is modded out.}
\label{fig:mirrTN}
\end{figure}
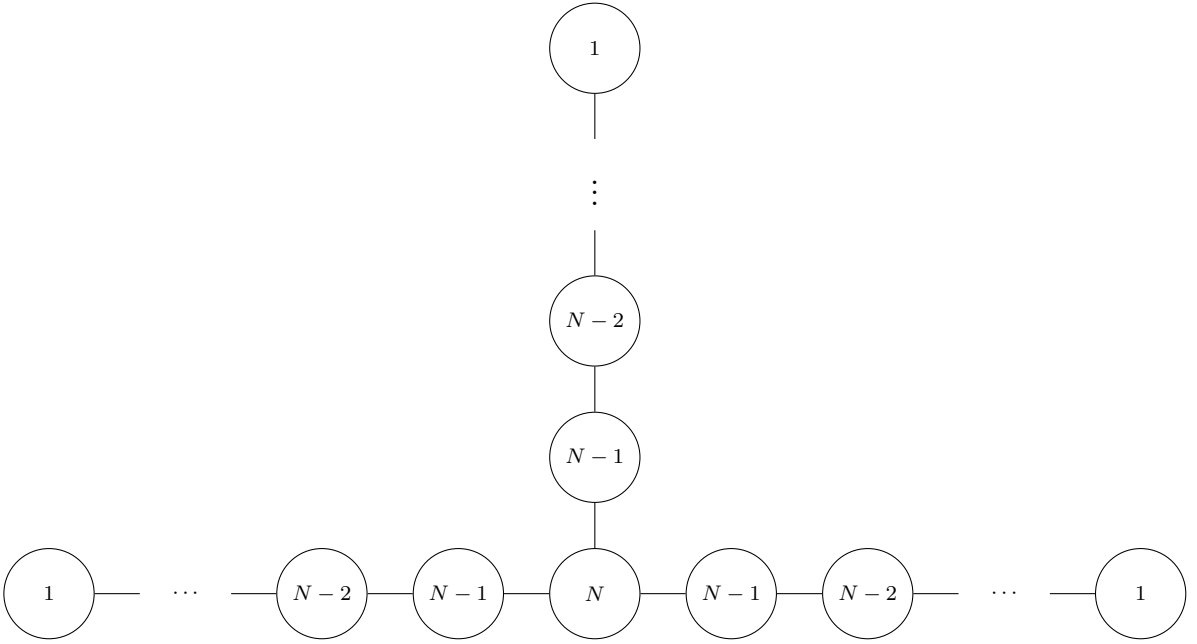

The case of a sphere with three maximal punctures $\vec\rho=(1,\cdots,1)$ is known as the $T_N$ theory \cite{Gaiotto:2009we}. 
We can compute the Coulomb branch Hilbert series of the mirror of the $T_N$ theory reduced to three dimensions by gluing three $T(SU(N))$ tails together. The quiver diagram of such a mirror theory is depicted in \fref{fig:mirrTN}.

Note that for $N=3$ the quiver of the mirror is the $E_6$ quiver and the result  should match with the Hilbert series of the reduced moduli space of 1 $E_6$ instanton on $\BC^2$.
\be \label{HLformulaTN}
\begin{split}
&H[\text{mirror $T_N$}] (t; {\vec x}^{(1)}, {\vec x}^{(2)}, {\vec x}^{(3)}) \\
&= \sum_{n_1 \geq \cdots \geq n_N =0} \left \{ \prod_{j=1}^3 H[{T (SU(N))}] (t; {\vec x}^{(j)} ; n_1,\ldots, n_{N}) \right \}\times \\
& \quad  t^{-\delta_{U(N)} (n_1,\ldots, n_{N})}  (1-t) P_{U(N)} (t; n_1,\ldots,n_N) \; \epsilon^{\sum_{i=1}^N n_i} \\
&=  \sum_{n_1 \geq \cdots \geq n_{N-1} \geq 0} t^{\frac{1}{2}\sum_{j=1}^{N-1} (N+1-2j) n_j} (1-t)^{3N+1} P_{U(N)} (t;n_1, \ldots, n_{N-1},0)  \times  \\
& \qquad  \epsilon^{\sum_{i=1}^{N-1} n_i}\prod_{j=1}^3 K_{(1^N)} ({\vec x}^{(j)};t)  \Psi_{U(N)}^{(n_1,\ldots, n_{N-1},0)}({\vec x}^{(j)};t)~,
\end{split}
\ee
where we explain the notation below:
\bi
\item 
${\vec x}^{(i)} =(x^{(i)}_1, \dots , x^{(i)}_N)$, with $i=1,2,3$, denotes the fugacities of the $SU(N)$ global symmetry on the Coulomb branch associated with the $i$-th copy of $T(SU(N))$; they satisfy 
\bea
\prod_{k=1}^N  x^{(i)}_k =1~, \qquad \text{with $i=1,2,3$}~.
\eea
\item The second line of the first equality is the gluing factor for the $U(N)$ group:
\ben
\item $\delta_{U(N)}$ denotes the contribution from the $U(N)$ background vector multiplet:
\be \label{vecUNnormalization}
\begin{split}
\delta_{U(N)}&(n_1,\ldots, n_{N}) = \sum_{1\leq i <j \leq N} |n_i-n_j|  \\
&= \sum_{j=1}^{N} (N+1-2j)n_j~, \qquad n_1 \geq n_2 \geq \cdots \geq n_N \geq 0~.
\end{split}
\ee
\item The removal of the overall $U(1)$ is done in two steps:
\ben
\item Multiplying $(1-t)$ to the function $P_{U(N)} (t; n_1,\dots,n_N)$.
\item Restricting $n_N=0$.  
\een
\een
\item The prefactor $K_{(1^N)} (\vec x; t)$ is given by
\bea
K_{(1^N)} (\vec x; t) &= \PE \left[ \chi^{U(N)}_{{\bf Adj}} (\vec x) t \right]~.
\eea
\item The fugacity $\epsilon$, with $\epsilon^N=1$, corresponds to a potential $\BZ_N$ \emph{discrete topological symmetry} for the $U(N)$ gauge group modulo $U(1)$. In the notations of section 3 of \cite{Cremonesi:2013lqa}, the $\bZ_N$ valued fugacity is related to the ambiguity in taking the $N$-th root when solving the constraint on the topological fugacities for $z_0$:
\bea\label{soln_constrnt_ambig}
z_0 = \epsilon \hat{z}_0~, \qquad \text{with} \quad \hat{z}_0 :=\left( \prod_{a=1}^e \prod_{k=1}^{d_a} z_{k,a}^{N_{k,a}}\right)^{1/N}~,
\eea
where $\hat{z}_0$ denotes the $N$-th principal root and $\epsilon$ runs over  $N$-th roots of unity,
\bea
\epsilon^N=1~, \label{epsilon}
\eea
and $N_{k,a}$ and $z_{k,a}$ are the rank and the fugacity for the topological symmetry of the $k$-th gauge group in the $a$-th leg. Often all or part of this  $\bZ_N$ symmetry can be absorbed in the center of the continuous topological symmetry associated to $z_{k,a}$. For this reason we will sometimes omit $\epsilon$ in the following.
\ei

Our result should agree with the Higgs branch Hilbert series of the  $T_N$ theory. The  latter can  be evaluated in the 4d version of the theory, since the Higgs branch does not depend on the dimension.
Let us compare \eref{HLformulaTN} with the result in \cite{Gadde:2011uv} for the Higgs branch Hilbert series of $T_N$ which is computed by the 4d Hall-Littlewood index. In that reference, the HL polynomial is defined with a normalization factor:%
\footnote{Our fugacity $t$ is related to $\tau$ in \cite{Gadde:2011uv} by $\tau=t^{1/2}$.}
\bea \label{normHLUN}
\widehat{\Psi}_{U(N)}^{\vec \lambda}(x_1,\dots,x_N;t)={\mathcal N}_{\vec \lambda}(t) \Psi_{U(N)}^{\vec \lambda}(x_1,\dots,x_N;t)~.
\eea
The normalization ${\mathcal N}_{\vec \lambda}(t)$ 
is given by
\be\label{normHL1}
{\mathcal N}^{-2}_{\lambda_1,...\lambda_k}(t)=\prod_{i=0}^\infty \prod_{j=1}^{m(i)}\,
 \left(\frac{1-t^{j}}{1-t}\right)\, ,
\ee 
where $m(i)$ is the number of rows in the Young diagram ${\vec \lambda}=(\lambda_1,\dots,\lambda_N)$ of length $i$. It
is related to $P_{U(N)}$ as follows:
\bea \label{iden1}
(1-t)^{N} P_{U(N)} (t; n_1, \ldots, n_{N-1},0) &= \CN_{n_1,\ldots, n_{N-1},0}(t)^2~.
\eea
Using the identity
\bea \label{iden2}
(1-t)^{2N+1} t^{\frac{1}{2}\sum_{j=1}^{N-1} (N+1-2j) n_j} &= \frac{(1-t)^{N+2}  \prod_{i=2}^N(1-t^i) }{\Psi_{U(N)}^{(n_1,\ldots, n_{N-1},0)} (t^{\frac{1}{2}(N-1)}, t^{\frac{1}{2}(N-3)}, \ldots, t^{-\frac{1}{2}(N-1)};t)}~,
\eea
we arrive at
\bea
&H[\text{mirror $T_N$}] (t; {\vec x}^{(1)}, {\vec x}^{(2)}, {\vec x}^{(3)})  \nn\\
& = (1-t)^{N+2} \left\{ \prod_{i=2}^N(1-t^i)\right\} K_{(1^N)}({\vec x}^{(1)};t)K_{(1^N)}({\vec x}^{(2)};t)K_{(1^N)}({\vec x}^{(3)};t) \times \nn \\
&  \sum_{n_1\geq n_2 \geq \cdots \geq n_{N-1} \geq 0} \frac{\widehat{\Psi}_{U(N)}^{(n_1,\ldots, n_{N-1},0)} ({\vec x}^{(1)};t)\widehat{\Psi}_{U(N)}^{(n_1,\ldots, n_{N-1},0)} ({\vec x}^{(2)};t)\widehat{\Psi}_{U(N)}^{(n_1,\ldots, n_{N-1},0)} ({\vec x}^{(3)};t) }{\widehat{\Psi}_{U(N)}^{(n_1,\ldots, n_{N-1},0)} (t^{\frac{1}{2}(N-1)}, t^{\frac{1}{2}(N-3)}, \ldots, t^{-\frac{1}{2}(N-1)};t)}~,
\eea
where the normalized HL polynomial $\widehat{\Psi}^{\vec n}_{U(N)}(\vec x;t)$ is defined as in \eref{normHLUN}.  Our result agrees with formula (5.33) of \cite{Gadde:2011uv}.

\subsection{The Coulomb branch of the mirror of a general $3d$ Sicilian theory}
The computation of the Coulomb branch Hilbert series for the mirror of $T_N$ can be easily generalized to a general $3d$ Sicilian theory.  For the mirror of a theory that arises from a compactification of the $A_{N-1}$ $6d$ $(2,0)$ theory on a circle times a genus $g$ Riemann surface with punctures $\{ \vec \rho_1, \vec \rho_2, \ldots, \vec \rho_e \}$, the Coulomb branch Hilbert series is given by
\be \label{HLformulaSicilian}
\begin{split}
H[\text{mirror $g$},&\, \{ \vec \rho_1, \vec \rho_2, \ldots, \vec \rho_e \}] (t; {\vec x}^{(1)}, \ldots, {\vec x}^{(e)}) \\
&= \sum_{n_1 \geq \cdots \geq n_N =0} \left \{ \prod_{j=1}^e H[{T_{\vec \rho_j}(SU(N))}] (t; {\vec x}^{(j)} ; n_1,\ldots, n_{N}) \right \}\times \\
& \qquad  t^{{\tilde \delta}_{U(N),\,g} (n_1,\ldots, n_{N})}(1-t) P_{U(N)} (t; n_1,\ldots,n_N) ,
\end{split}
\ee
where the contribution of the $g$ $U(N)$ adjoint hypermultiplets and vector multiplet to the dimension of monopole operators is
\be
\begin{split} \label{vecUN}
\tilde {\delta}_{U(N),\,g} (\vec n) &= (g-1) \delta_{U(N)}(\vec n) =(g-1) \sum_{1 \leq i < j \leq N} |n_i-n_j|  \nn \\
&= (g-1) \sum_{j=1}^{N} (N+1-2j)n_j~, \qquad n_1 \geq \cdots \geq n_N \geq 0~,
\end{split}
\ee
with $\delta_{U(N)}(n_1,\ldots, n_{N})$ given by \eref{vecUN}.
We therefore obtain
\be \label{genSicilian}
\begin{split}
&H[\text{mirror $g, \{ \vec \rho_1, \vec \rho_2, \ldots, \vec \rho_e \}$}] (t; {\vec x}^{(1)}, \ldots, {\vec x}^{(e)}) \\
&=  \sum_{n_1 \geq \cdots \geq n_{N-1} \geq 0} t^{(\frac{e}{2}+g -1)\sum_{j=1}^{N-1} (N+1-2j) n_j} (1-t)^{eN+1} P_{U(N)} (t;n_1, \ldots, n_{N-1},0) \times \\
& \qquad \prod_{j=1}^e K_{\vec \rho_j} ({\vec x}^{(j)};t)  \Psi_{U(N)}^{(n_1,\ldots, n_{N-1},0)}({\vec x}^{(j)} t^{\frac{1}{2} \vec w_{\rho_j}};t)~,
\end{split}
\ee

\subsubsection{The case of genus zero} 
In a special case of $g=0$, we use \eref{iden1} and \eref{iden2} to obtain
\be \label{genus0}
\begin{split}
&H[\text{mirror $\{ \vec \rho_1, \vec \rho_2, \ldots, \vec \rho_e \}$}](t; {\vec x}^{(1)}, \ldots, {\vec x}^{(e)})  \\
& = (1-t)^{e+(N-1)} \left\{ \prod_{i=2}^N(1-t^i)\right\}^{e-2}  \times \\
& \qquad  \sum_{n_1\geq n_2 \geq \cdots \geq n_{N-1} \geq 0} \frac{ \prod_{j=1}^e K_{\vec \rho_j}({\vec x}^{(j)};t) \widehat{\Psi}_{U(N)}^{(n_1,\ldots, n_{N-1},0)} ({\vec x}^{(j)} t^{\frac{1}{2} \vec w_{\rho_j}};t)}{[\widehat{\Psi}_{U(N)}^{(n_1,\ldots, n_{N-1},0)} (t^{\frac{1}{2}(N-1)}, t^{\frac{1}{2}(N-3)}, \ldots, t^{-\frac{1}{2}(N-1)};t)]^{e-2}}~,
\end{split}
\ee
where $\widehat{\Psi}_{U(N)}$ denotes the normalized Hall-Littlewood polynomial defined in \eref{normHLUN}.
This result agrees with the Higgs branch Hilbert series of the Gaiotto theory, computed as a Hall-Littlewood index for $g=0$ in (2.13) of \cite{Gaiotto:2012uq}. 

\begin{figure}
\begin{center}
\includegraphics[scale=0.35]{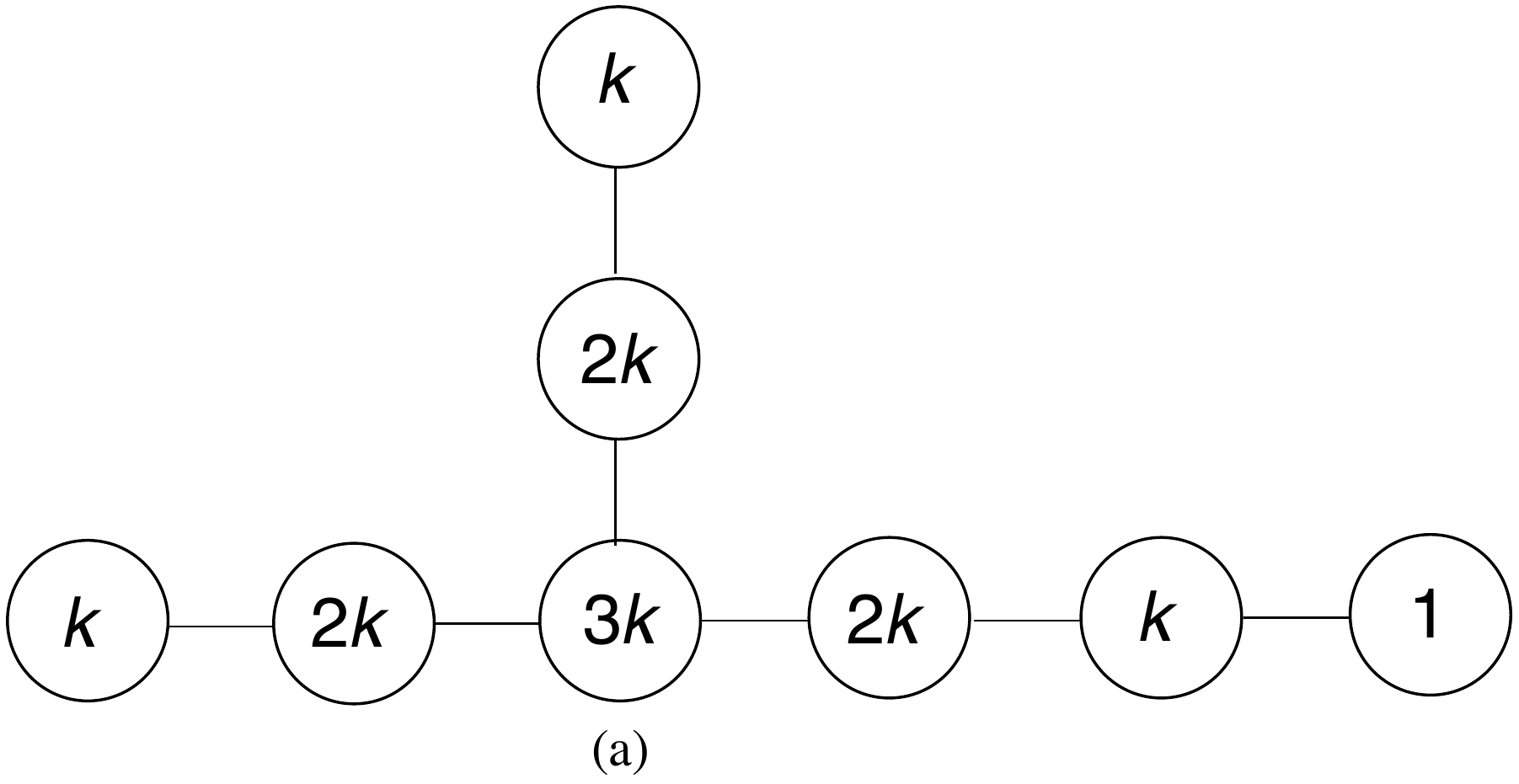} \qquad  \qquad
\includegraphics[scale=0.35]{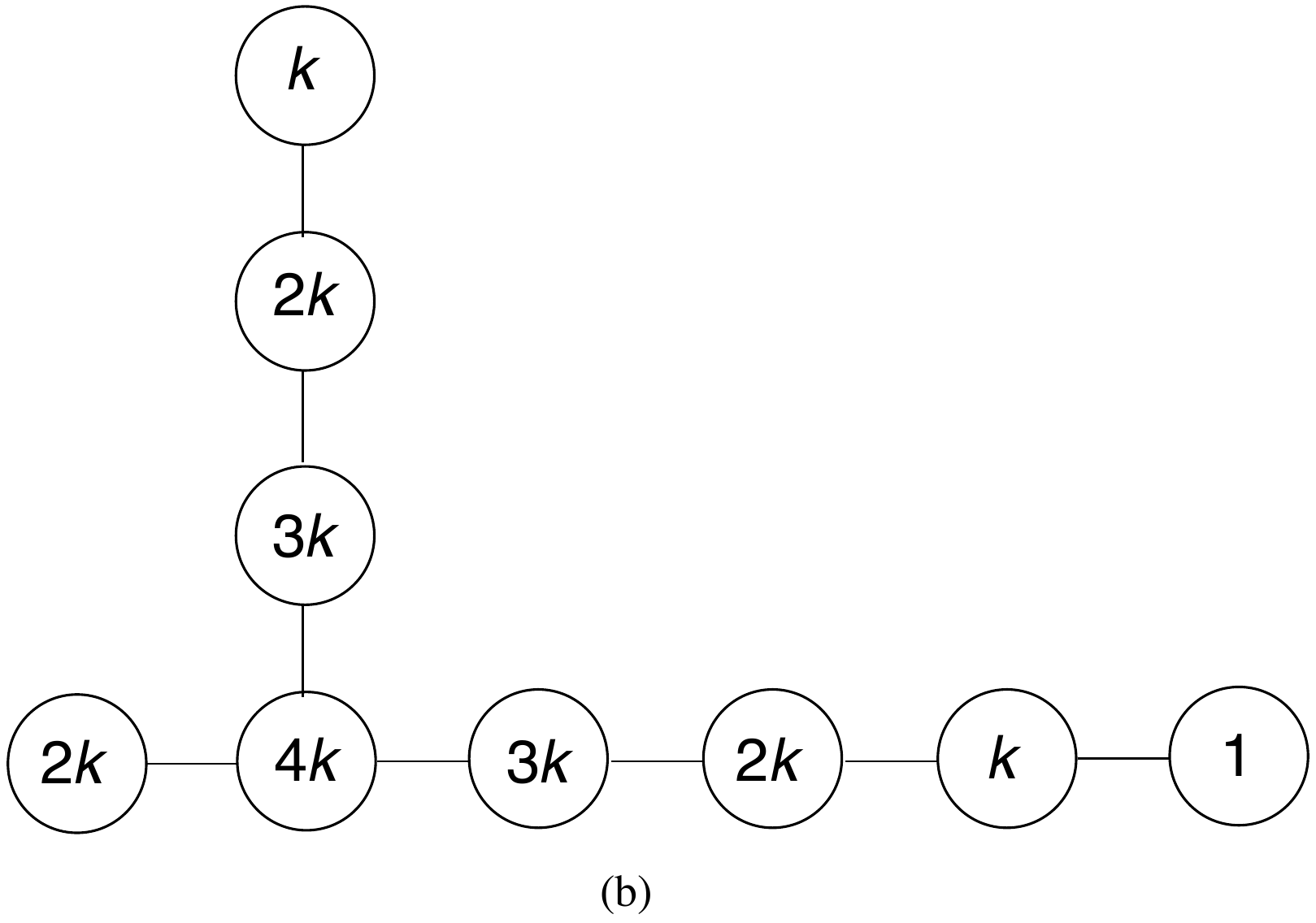} \quad \\~\\
\includegraphics[scale=0.40]{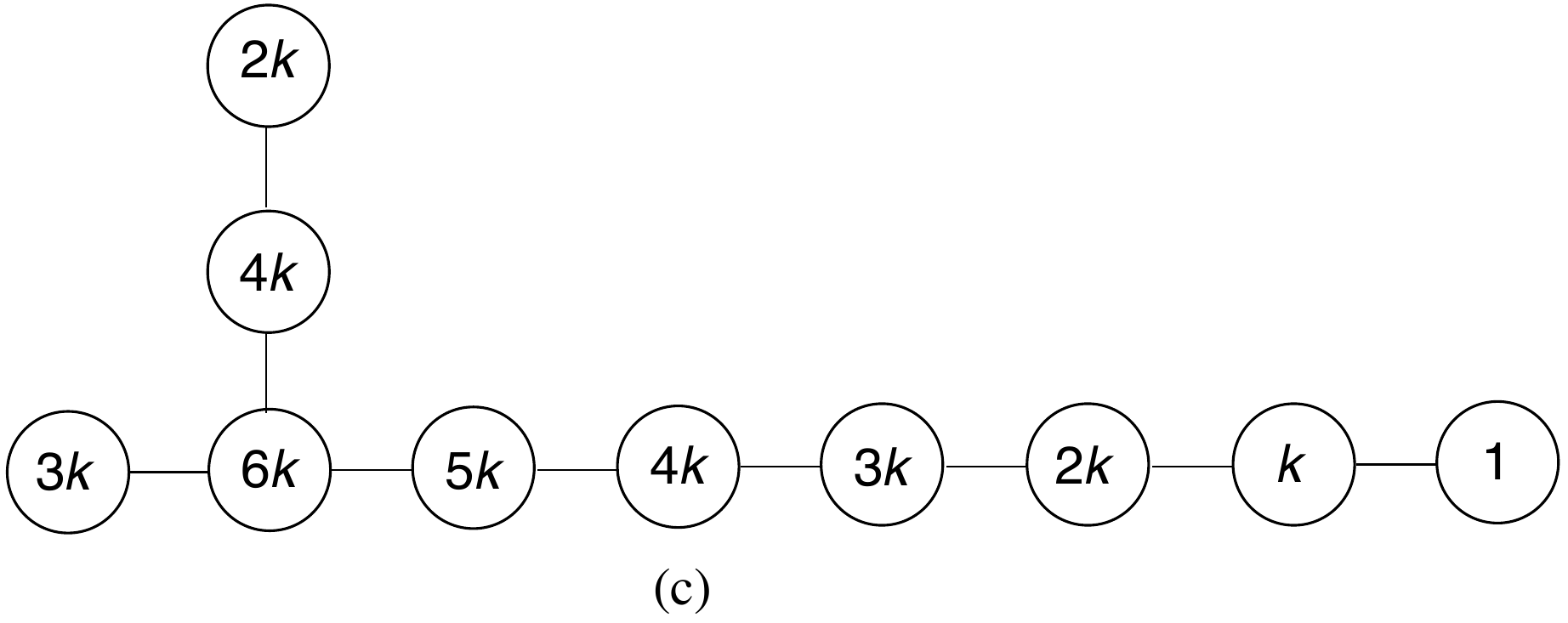} \quad 
\caption{The moduli spaces of $k$ $E_6$, $E_7$ and $E_8$ instantons on $\BC^2$ can be realized using the Coulomb branch of quiver diagrams (a), (b) and (c) respectively.  Each node represents a unitary group of the labelled rank and the overall $U(1)$ is modded out in each diagram.}
\label{fig:E678}
\end{center}
\end{figure}

As discussed in \cite{Gaiotto:2012uq},  the  formula \eref{genus0} can be used to write the Hilbert series for the moduli spaces of $E_6$, $E_7$ and $E_8$ instantons on $\BC^2$, which can be realized as  the Higgs branch of the $6d$ $(2,0)$ theory compactified on a Riemann sphere with punctures $\{\vec \rho_1, \vec \rho_2, \vec \rho_3\}$
\bea
\begin{array}{llll}
~ &\qquad \vec \rho_1 & \qquad \vec \rho_2 & \qquad  \quad \vec \rho_3 \\
E_6 &\quad (k,k,k) &\quad (k,k,k) &\quad (k,k,k-1,1) \\
E_7 &\quad (k,k,k,k) &\quad (2k,2k) &\quad (k,k,k,k-1,1) \\
E_8 &\quad (3k,3k) &\quad (2k,2k,2k) & \quad (k,k,k,k,k,k-1,1)
\end{array}
\eea
 corresponding to the mirror quiver given in Figure \ref{fig:E678}. 

\subsubsection{Mirror of the $SU(3)$ Sicilian theory with $g=1$ and a maximal puncture} 
 Recall that for genus $g>0$ the HL index differs from the Higgs branch Hilbert series of the Sicilian theory \cite{Gadde:2011uv}. 
The latter is given by our formula \eqref{genSicilian}, assuming mirror symmetry.

\begin{figure}[H]
\begin{center}
\includegraphics[scale=0.4]{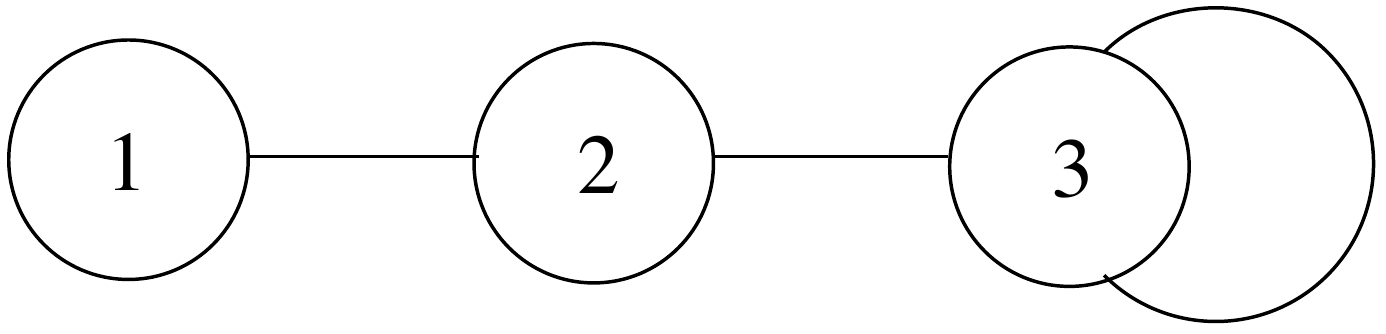}
\caption{Quiver for the mirror of the $A_2$ theory on a circle times a torus with one maximal puncture. 
The overall $U(1)$ is factored out.}
\label{fig:G2enh}
\end{center}
\end{figure}

Let us provide an explicit example for the case of $N=3$, $g=1$ and one maximal puncture $\vec \rho=(1,1,1)$ below.  The quiver diagram of the mirror theory of our interest is depicted in \fref{fig:G2enh}. This example is particularly interesting because the global symmetry on the Coulomb branch enhances to $G_2$ \cite{Gaiotto:2012uq}. 
We will show this by computing the Hilbert series and expanding it in $G_2$ characters.

The Coulomb branch Hilbert series can be computed using \eref{genSicilian}, where the fugacities $x_1, x_2, x_3$ are related to the fugacities for the topological charges of $U(1)$, $U(2)$ and $U(3)$ gauge groups and are subject to the constraint \eqref{constrx}.
In order to make $G_2$ characters manifest in the Hilbert series, we use the fugacity map%
\footnote{Here we use the characters of $G_2$ as in {\tt LiE} online service at the following link: {\url{http://young.sp2mi.univ-poitiers.fr/cgi-bin/form-prep/marc/LiE_form.act?action=character&type=G&rank=2&highest_rank=8}}.}
\bea
x_1 = y_1, \qquad x_2 = y_1 y_2^{-1}~, \qquad x_3 =  y_1^{-2} y_2~,
\eea
where $x_1, x_2$ are the fugacities in formula \eref{genSicilian} and $y_1, \; y_2$ are the $G_2$ fugacities.  

We then obtain
\bea\label{G2char_exp}
H[\text{mirror $g=1, (1,1,1)$}] (t;y_1, y_1 y_2^{-1}, y_2 y_1^{-2}) = f(0,0,0)+f(3,1,5)~,
\eea
where  
\bea
f(a,b,c) = \sum_{n_1=0}^\infty \sum_{n_2=0}^\infty \sum_{n_3=0}^\infty \sum_{n_4=0}^\infty [2 n_2 + 3 n_3+a, n_1 + 2 n_4+b] t^{n_1 + 2 n_2 + 3 n_3 + 4 n_4+c}~,
\eea
and $[a,b]$ denotes the character of the $G_2$ representation with highest weight $[a,b]$, written in terms of $y_1, \; y_2$.
The character expansion \eqref{G2char_exp} shows not only that the adjoint representation arises at $\Delta=1$ (for the scalar partners of conserved currents), but also that the whole chiral spectrum transforms in $G_2$ representations as expected.

The unrefined Hilbert series is given by
\bea
H[\text{mirror $g=1, (1,1,1)$}] (t;1,1,1) = \frac{1 + 4 t + 9 t^2 + 9 t^3 + 4 t^4 + t^5}{(1 - t)^{10}}~,
\eea
with a palindromic numerator and a pole at $t=1$ of order $10$, equal to the complex dimension of the Coulomb branch of the moduli space.

\paragraph{The generating function of highest weights \cite{Hanany:2014dia}.} The highest weight vectors that appear in formula \eref{G2char_exp} can be collected in the following generating function:
\be \label{Hanany:2014diaen}
\begin{split}
&\PE [ \mu_2 t + \mu_1^2 t^2 + \mu_1^3 t^3 + \mu_2^2 t^4 + \mu_1^3 \mu_2 t^5 - \mu_1^6 \mu_2^2 t^{10}] \\
&= \frac{1-t^{10} \mu _1^6 \mu _2^2}{\left(1-t^2 \mu _1^2\right) \left(1-t^3 \mu _1^3\right) \left(1-t \mu _2\right) \left(1-t^5 \mu _1^3 \mu _2\right) \left(1-t^4 \mu _2^2\right)}~,
\end{split}
\ee
where $\mu_1$ and $\mu_2$ are the fugacities associated with the highest weights $n_1$ and $n_2$ of representations of $G_2$.  Upon computing the power series in $t$ of \eref{Hanany:2014diaen}, the powers $\mu_1^{n_1}\mu_2^{n_2}$ 
can be traded for the Dynkin label $[n_1,n_2]$ to obtain the character expansion as stated in \eref{G2char_exp}.  Let us demonstrate this for the first few terms in the power series:
\bea
1+\mu _2 t+\left(\mu _1^2+\mu _2^2\right) t^2+\left(\mu _1^3+\mu _1^2 \mu _2+\mu _2^3\right) t^3+\ldots~.
\eea
Trading the powers of $\mu_1$ and $\mu_2$ for the Dynkin label, we obtain
\bea
1+[0,1] t+([2,0]+[0,2]) t^2 +([3,0]+[2,1]+[0,3])t^3 +\ldots~.
\eea

\section{Mirrors of $3d$ Sicilian theories of $D$-type} \label{sec:SicilianD}

In this section we consider three dimensional theories arising from the $6d$ $(2,0)$ theory of $D_N$ type compactified on a circle times a Riemann surface with punctures.  Each puncture is classified by a $D$-partition of $SO(2N)$. The Coulomb branch Hilbert series of the mirror theory can be computed by gluing copies of the $T_{\vec \rho}(SO(2N))$ theories \cite{Benini:2010uu} according to the general discussion in section \ref{sec:Sicilian}. The quivers for the $T_{\vec \rho}(SO(2N))$ theories, which can be realised from brane and orientifold configurations as in \cite{Feng:2000eq}, are reviewed in section 4.2 of \cite{Cremonesi:2014kwa}. 
We remark that we gauge the centerless group $SO(2N)/\bZ_2$ rather than $SO(2N)$. Consequently, the magnetic fluxes of the gluing gauge group belong to the weight lattice of the dual group $Spin(2N)$ modulo the Weyl group. 

Given a $3d$ Sicilian theory with genus $g$ and $e$ $D$-type punctures $ \{ \vec \rho_1, \vec \rho_2, \ldots, \vec \rho_e \}$, the Coulomb branch Hilbert series of its mirror theory is%
\footnote{It is straightforward to include in \eref{Sicilian_D} a fugacity for the center of $Spin(2N)$, but we prefer not to clutter formulae with those factors, which can often be reabsorbed. }
\be\label{Sicilian_D} 
\begin{split}
H[\text{mirror $g$},&\, \{ \vec \rho_1, \vec \rho_2, \ldots, \vec \rho_e \}] (t; {\vec x}^{(1)}, \ldots, {\vec x}^{(e)}) \\
&= \sum_{n_1 \geq \cdots \geq n_{N-1}\ge |n_N |} \left \{ \prod_{j=1}^e H[{T_{\vec \rho_j}(SO(2N))}] (t; {\vec x}^{(j)} ; n_1,\ldots, n_{N}) \right \}\times \\
& \qquad  t^{{\tilde \delta}_{SO(2N),\,g} (n_1,\ldots, n_{N})}P_{SO(2N)} (t; n_1,\ldots,n_N) ,
\end{split}
\ee
where $H[{T_{\vec \rho}(SO(2N))}]$ is given by \eref{mainHL}, the Casimir factor $P_{SO(2N)}$ is computed as in \eqref{classical_dressing} (see (A.10) of \cite{Cremonesi:2013lqa} for an explicit expression), and ${\tilde \delta}_{SO(2N),\,g} (\vec n)$ is the contribution of the $g$ $SO(2N)$ adjoint hypermultiplets and vector multiplet to the dimension of monopole operators is
\bea
{\tilde \delta}_{SO(2N),\,g} (\vec n) = (g-1) \delta_{SO(2N)} (\vec n) = (g-1)\sum_{j=1}^{N-1} (2N-2j)n_j~,
\eea
with the second equality following from \eref{powers}. Note that because the dual of the gluing group is $Spin(2N)$, $n_1, \dots, n_{N}$ are {\bf all integers or all half-odd integers}. 

For $g=0$ our formula \eref{Sicilian_D} for the Coulomb branch Hilbert series of mirrors of D-type Sicilian theories proposed in \cite{Benini:2010uu} agrees with the Higgs branch Hilbert series of the Sicilian theory, computed as the Hall-Littlewood limit of the superconformal index of the 4d Sicilian theory in formula (4.10) of \cite{Lemos:2012ph}.%
\footnote{The orthonormal Hall-Littlewood polynomials 
used in \cite{Lemos:2012ph} can be expressed in terms of the Hall-Littlewood polynomials used here as $P_{M~ G}^{\vec n}(\vec{a}|0,t)=(1-t)^{rk(G)/2} P_{G^\vee}(t;\vec n)^{1/2} \Psi^{\vec n}_G(\vec{a}(t,{\vec x});t)$. The pre-factors are related by $\cK_G=(1-t)^{rk(G)/2} K^G$. Finally, for $G=SO(2N)$ one finds $\cA(0,t)/P_{M~ SO(2N)}^{\vec n}(1,t,t^2,\dots,t^{N-1}|0,t) = t^{\frac{1}{2}\delta_{SO(2N)}(\vec{n})} P_{SO(2N)}(t;\vec n)^{-1/2}$.}
For higher genus the HL index does not compute the Hilbert series of the Higgs branch. Formula \eref{Sicilian_D} provides a prediction for the latter, assuming mirror symmetry.

In the rest of the section we provide examples of Sicilian theories with $D_3$ and $D_4$ symmetry and we compare with the results in \cite{Lemos:2012ph,Chacaltana:2011ze}. We start this section by considering the case of $D_3$.  Due to the isomorphism of its Lie algebra with that of $A_3$, each $D_3$ puncture can be identified with an $A_3$ puncture.
We compute the Coulomb branch Hilbert series of mirror theories of $3d$ Sicilian theories with $D_3$ punctures using the Hall-Littlewood formula and compare the result with those with $A_3$ punctures. We then consider $D_4$ theories with a set of punctures for which the Higgs branch is explicitly known and we compare our result for the Coulomb branch Hilbert series of the mirror with  the Higgs branch Hilbert series. The case of twisted D punctures is discussed in the Appendix. All these examples demonstrate the validity of our formula \eref{Sicilian_D}.

\subsection{$D_3$ punctures}

There are four possible $D$-partitions of $SO(6)$.  These partitions and the identification with $A_3$ partitions are given on Page 17 of \cite{Chacaltana:2011ze}.  We list them as follows in \tref{tab:D3punc}.
\begin{table}[H]
\begin{center}
\begin{tabular}{|c|c|c|}
\hline
$D_3$ puncture & $A_3$ puncture & Global symmetry\\
\hline
$(1^6)$ & $(1^4)$ & $SO(6) \simeq SU(4)$ \\
\hline
$(2^2, 1^2)$ & $(2,1^2)$ & $USp(2) \times SO(2) \simeq SU(2) \times U(1)$ \\
\hline
$(3, 1^3)$ & $(2^2)$ & $SO(3) \simeq SU(2)$ \\
\hline
$(3^2)$ & $(3,1)$ & $SO(2) \simeq U(1)$ \\
\hline 
\end{tabular}
\caption{The list of $D_3$ regular punctures, their identifications with $A_3$ punctures and the associated global symmetries.}
\label{tab:D3punc}
\end{center}
\end{table}%
Next, we consider an example of the mirror theory of a $3d$ Sicilian theory with $D_3$ punctures $(3^2)$, $(1^6)$ and $(1^6)$.

\subsubsection{$D_3$ punctures: $(3^2)$, $(1^6)$ and $(1^6)$} 
In terms of $A_3$ punctures, these punctures correspond to two maximal $(1^4)$ and one minimal $(3,1)$ punctures.  This Sicilian theory corresponds to the quiver diagram $[SU(4)]-[SU(4)]$, and contains 16 free hypermultiplets; see \cite{Gaiotto:2009we} and page 18 of \cite{Chacaltana:2010ks}.

The Coulomb branch Hilbert series of the mirror theory of this Sicilian theory can be computed by gluing two copies of $T_{(1^6)} (SO(6))$ and one copy of $T_{(3^2)} (SO(6))$ together via the common $SO(6)$ symmetry:
\bea \label{free16}
 H(t; \vec x, \vec y,z) &= \sum_{a_1, a_2, a_3 \geq 0} t^{-\delta_{SO(6)}(\vec n(\vec a))} P_{SO(6)} (t;\vec n(\vec a)) \; H[T_{(1^6)} (SO(6))] (t; \vec x; \vec n(\vec a))\times \nn \\
& \qquad H[T_{(1^6)} (SO(6))] (t;\vec y;\vec n(\vec a)) H[T_{(3^2)} (SO(6))] (t; z; \vec n(\vec a))~,
\eea
where $\vec x, \vec y, z$ are respectively fugacities of $SO(6)$, $SO(6)$ and $SO(2)$ symmetries and the function $P_{SO(6)}$ is defined as in (A.10) of \cite{Cremonesi:2013lqa}, and
\bea
\vec n( \vec a) &= \left( a_1+\frac{1}{2}(a_2+a_3), \frac{1}{2}(a_2+a_3), \frac{1}{2}(-a_2+a_3) \right)~, \nn \\
\delta_{SO(6)} (\vec n) &= 4n_1 + 2n_2~,  \\
H[T_{(1^6)} (SO(6))] (t; \vec x; \vec n) &= t^{\frac{1}{2}\delta_{SO(6)} (\vec n)} (1-t)^3 \PE\left[ t \chi^{D_3}_{[0,1,1]} (\vec x) \right] \Psi_{D_3} (\vec x; \vec n; t)~, \nn \\
H[T_{(3^2)} (SO(6))] (t; x; \vec n) &= t^{\frac{1}{2}\delta_{SO(6)} (\vec n)} (1-t)^3 \PE\left[ t +t^2 \chi^{C_1}_{[2]}(x) +t^3\right] \Psi_{D_3} (t x, t^{-1} x, x; \vec n; t)~. \nn
\eea
Note that in the above notation, $\vec a= [a_1, a_2, a_3]$ denotes a Dynkin label of an irreducible representation of $Spin(6)$ and $\vec n =(n_1, n_2, n_3)$ denotes its highest weight in the standard basis.  Hence  the summations run over all irreducible representations of $Spin(6)$, including the spinorial representations.

It can be checked that the first few terms in the power series of \eref{free16} are equal to those of the Hilbert series of $16$ free hypermultiplets in the spinor representations of $SO(6)$, as expected from mirror symmetry:
\bea
H(t; \vec x, \vec y,z) &= \PE \left[ \left \{ z^{1/2} \chi^{D_3}_{[0,1,0]}(\vec x)  \chi^{D_3}_{[0,1,0]}(\vec y)  + z^{-1/2} \chi^{D_3}_{[0,0,1]}(\vec x)  \chi^{D_3}_{[0,0,1]}(\vec y) \right \} t \right] \nn \\
&= \sum_{n_1, n_2,n_3 =0}^\infty \chi^{D_3}_{[n_2, n_1,n_3]}(\vec x) \chi^{D_3}_{[n_2, n_1,n_3]}(\vec y) (z^{1/2} t)^{n_1+2n_2+3n_3}  \times \nn \\
& \qquad \sum_{m_1,m_2,m_3 =0}^\infty \chi^{D_3}_{[m_2, m_3,m_1]}(\vec x) \chi^{D_3}_{[m_2, m_3,m_1]}(\vec y) (z^{-1/2} t)^{m_1+2m_2+3m_3} ~.
\eea

\subsection{$D_4$ punctures}
In this section, we provide three examples on Sicilian theories with the following $D_4$ punctures.
\ben
\item $(5,3)$, $(2^2, 1^4)$ and $(1^8)$~,
\item $(3^2, 1^2)$, $(2^2, 1^4)$ and $(2^2, 1^4)$~,
\item $(5,3)$, $(5,3)$, $(2^4)$ and $(3,1^5)$~.
\een
In the following subsections, we compute the Coulomb branch Hilbert series of the mirror theories of these Sicilian theories and compare the results to those presented in \cite{Chacaltana:2011ze}.

For reference, we tabulate the quiver diagrams for $T_{\vec \rho}(SO(8)$, with $\vec \rho$ being partitions listed above, in \tref{tab:quivexamples}.
\begin{table}[htdp]
\begin{center}
\begin{tabular}{|l|l|}
\hline
Partition $\vec \rho$ & Quiver diagram for $T_{\vec \rho} (SO(8))$ \\
\hline
$(5,3)$ & $[SO(8)]-(USp(2))$\\
$(3^2, 1^2)$ & $[SO(8)]-(USp(4))-(SO(2))$\\
$(2^4)$ & $[SO(8)]-(USp(6))-(SO(4))-(USp(2))$\\
$(3,1^5)$ & $[SO(8)]-(USp(4))-(SO(4))-(USp(2))-(SO(2))$ \\
$(2^2,1^4)$ & $[SO(8)]-(USp(6))-(SO(4))-(USp(2))-(SO(2))$ \\
$(1^8)$ & $[SO(8)]-(USp(6))-(SO(6))-(USp(4))-(SO(4))-(USp(2))-(SO(2))$\\
\hline
\end{tabular}
\end{center}
\caption{Quiver diagrams for $T_{\vec \rho} (SO(8))$ for certain $D_4$ partitions $\vec \rho$.}
\label{tab:quivexamples}
\end{table}%

\subsubsection{$D_4$ punctures: $(5,3)$, $(2^2, 1^4)$ and $(1^8)$}
The global symmetries associated with these punctures are trivial, $USp(2) \times SO(4) \simeq SU(2)^3$ and $SO(8)$, respectively.
According to page 24 of \cite{Chacaltana:2011ze}, this Sicilian theory is a free theory containing 48 half-hypermultiplets.  

We realize the Higgs branch of this theory from the Coulomb branch of the mirror theory.  The quiver diagram of the latter can be obtained by gluing the quiver diagrams of $T_{(5,3)} (SO(8))$, $T_{(2^2, 1^4)} (SO(8))$ and $T_{(1^8)} (SO(8))$ via the common symmetry $SO(8)/\bZ_2$; this is depicted in \eref{quivex1}, where each gray node labeled by $N$ denotes an $SO(N)$ gauge group (with the central node $8*$ being $SO(8)/\BZ_2$) and each black node labeled by $M$ denotes a $USp(M)$ gauge group.
{\large
\bea \label{quivex1}
\node{}{2}-\Node{}{2}-\node{}{4}-\Node{}{6}-\node{\Ver{}{2}}{8*}-\Node{}{6}-\node{}{6}-\Node{}{4} -\node{}{4} -\Node{}{2} -\node{}{2}  \ \tikz[na]\node(B1){};
\eea}
Note that the ranks of all gauge groups add up to $24$. This is the quaternionic dimension of the Coulomb branch, which indeed agrees with the dimension of the Higgs branch of the theory of $48$ free half-hypermultiplets.

The Coulomb branch Hilbert series of the mirror theory is
\bea \label{free24}
 H(t; \vec x, \vec y) &= \sum_{a_1, a_2, a_3,a_4 \geq 0} t^{-\delta_{SO(8)}(\vec n(\vec a))} P_{SO(8)} (t;\vec n(\vec a)) \; H[T_{(5,3)} (SO(8))] (t; \vec n(\vec a))\times \nn \\
& \qquad H[T_{(2^2, 1^4)} (SO(8))] (t;\vec y;\vec n(\vec a)) H[T_{(1^8)} (SO(8))] (t; \vec x; \vec n(\vec a))~,
\eea
where $\vec x=(x_1,x_2,x_3)$ and $\vec y=(y_1,\ldots, y_4)$ are respectively fugacities of $SU(2)^3$ and $SO(8)$ symmetries and the function $P_{SO(8)}$ is defined as in (A.10) of \cite{Cremonesi:2013lqa}, and
\bea
\vec n( \vec a) &= \left( a_1+a_2+\frac{a_3+a_4}{2}, a_2+\frac{a_3+a_4}{2}, \frac{a_3+a_4}{2}, \frac{-a_3+a_4}{2} \right)~, \nn \\
\delta_{SO(8)} (\vec n) &= 6n_1+4n_2 + 2n_3~, \nn \\
H[T_{(1^8)} (SO(8))] (t; \vec x; \vec n) &= t^{\frac{1}{2}\delta_{SO(8)} (\vec n)} (1-t)^4 K_{(1^8)} (\vec x;t) \Psi_{D_4} (\vec x;\vec n; t)~,\nn \\
H[T_{(5,3)} (SO(8))] (t; \vec n) &= t^{\frac{1}{2}\delta_{SO(8)} (\vec n)} (1-t)^4 K_{(5,3)} (t) \Psi_{D_4} (1, t, t^{-1}, t^2;\vec n; t)~, \nn \\
H[T_{(2^2, 1^4)} (SO(8))] (t; \vec y; \vec n) &= t^{\frac{1}{2}\delta_{SO(8)} (\vec n)} (1-t)^4 K_{(2^2, 1^4)} (\vec y; t) \Psi_{D_4} (t x_1^{-1}, t y_1, y_2, y_3; \vec n; t)~, \nn \\
K_{(1^8)}(\vec x;t) &= \PE \left[ \chi^{D_4}_{[0,1,0,0]} (\vec x) t \right]~, \nn\\
K_{(5,3)} (t)  &= \PE \left[ 3t^2+t^3 +2t^4 \right]~, \nn \\
K_{(2^2, 1^4)}(t; \vec y) &= \PE \Big[ t\left(2+ \chi^{SU(2)}_{[2]}(y_1)+\chi^{SU(2)}_{[2]}(y_2)\chi^{SU(2)}_{[2]}(y_3) \right) \nn \\
& \qquad + t^{3/2}  \chi^{SU(2)}_{[2]}(y_1) \{ \chi^{SU(2)}_{[2]}(y_2)+ \chi^{SU(2)}_{[2]}(y_3) \} + t^2\Big]~.
\eea

It can be checked that the first few terms in the power series of \eref{free24} agrees with
\be
\begin{split}
H(t; \vec x, \vec y) &= \PE \Big[ \Big \{ \chi^{D_4}_{[1,0,0,0]}(\vec x)  \chi^{SU(2)}_{[1]}( y_1)  + \chi^{D_4}_{[0,0,1,0]}(\vec x)  \chi^{SU(2)}_{[1]}( y_2)   \\
& \qquad + \chi^{D_4}_{[0,0,0,1]}(\vec x)  \chi^{SU(2)}_{[1]}( y_3)  \Big \} t \Big]~,
\end{split}
\ee
namely the Hilbert series of $48$ free half-hypermultiplets, as expected from mirror symmetry.

\subsubsection{$D_4$ punctures: $(3^2, 1^2)$, $(2^2, 1^4)$ and $(2^2, 1^4)$}
 The quiver diagram of the mirror of this Sicilian theory can be obtained by gluing the quiver diagrams of $T_{(3^2, 1^2)} (SO(8))$, $T_{(2^2, 1^4)} (SO(8))$ and $T_{(2^2, 1^4)} (SO(8))$ via the common symmetry $SO(8)/\bZ_2$; this is depicted in \eref{quivex2}, where each gray node labeled by $N$ denotes an $SO(N)$ gauge group (with the central node $8*$ being $SO(8)/\BZ_2$) and each black node labeled by $M$ denotes a $USp(M)$ gauge group.
{\large
\bea \label{quivex2}
\node{}{2}-\Node{}{2}-\node{}{4}-\Node{}{6}-\node{\overset{\ver{}{2}}{\Ver{}{4}}}{8*}-\Node{}{6}-\node{}{4}-\Node{}{2} -\node{}{2}  \ \tikz[na]\node(B1){};
\eea}
The quaternionic dimension of the Coulomb branch of this theory, equal to the sum of the ranks of all gauge groups, is $21$.

The global symmetries associated with these punctures are respectively $SO(2)^2$, $SU(2)^3$ and $SU(2)^3$.  According to page 28 of \cite{Chacaltana:2011ze}, this Sicilian theory can be identified with the $T_4$ theory and the global symmetry enhances to $SU(4)^3$.  Indeed, the Higgs branch of the $T_4$ theory is 21 quaternionic dimensional; this is in agreement with the dimension of the Coulomb branch of the mirror theory.
 
The Coulomb branch Hilbert series of theory depicted in \eref{quivex2} is
\bea \label{T4inD4}
 H(t; \vec x, \vec y, \vec z) &= \sum_{a_1, a_2, a_3,a_4 \geq 0} t^{-\delta_{SO(8)}(\vec n(\vec a))} P_{SO(8)} (t;\vec n(\vec a)) \; H[T_{(3^2, 1^2)} (SO(8))] (t; \vec z; \vec n(\vec a))\times \nn \\
& \qquad H[T_{(2^2, 1^4)} (SO(8))] (t;\vec y;\vec n(\vec a)) H[T_{(2^2, 1^4)} (SO(8))] (t;\vec x;\vec n(\vec a))~,
\eea
where $\vec x = (x_1, x_2, x_3)$ and $\vec y =(y_1,y_2,y_3)$ are fugacities for $SU(2)^3$,  $\vec z =(z_1, z_2)$ are fugacities for $SO(2)^2$, and
\bea
H[T_{(3^2, 1^2)} (SO(8))] (t; \vec z; \vec n) &= t^{\frac{1}{2}\delta_{SO(8)} (\vec n)} (1-t)^4 K_{(3^2, 1^2)} (\vec z;t) \Psi_{D_4} (z_1 t, z_1 t^{-1}, z_1, z_2;\vec n; t)~, \nn \\
K_{(3^2, 1^2)} (\vec z;  t) &= \PE \left[ 2t + \left(z_1^2+1+z_1^{-2} + \sum_{\epsilon_1, \epsilon_2 = \pm 1}z_1^{\epsilon_1} z_2^{\epsilon_2} \right) t^2 +t^3\right]~.
\eea

Computing the power series in $t$ of the above expression \eref{T4inD4}, we find that at order $t$, the $45$ gauge invariants transform as follows:
\be
\begin{split}
&(z_1+z_1^{-1}) [1]_{x_1}[1]_{y_1} + [2]_{x_1} +[2]_{y_1} +1 \\
&+(z_1^{1/2} z_2^{1/2} + z_1^{-1/2} z_2^{-1/2}) [1]_{x_2}[1]_{y_2} +[2]_{x_2}+ [2]_{y_2} +1  \\
&+(z_1^{1/2} z_2^{-1/2} + z_1^{-1/2} z_2^{1/2}) [1]_{x_3}[1]_{y_3} +[2]_{x_3}+ [2]_{y_3} +1 ~,
\end{split}
\ee
where $[\cdots]_{\vec a}$ denotes the character of representation $[\cdots]$ written in terms of $\vec a$.  Note that each line gives the decomposition of the adjoint representation of $SU(4)$ in terms of representations of $SO(2) \times SU(2)^2$. Hence these $45$ generators indeed decompose into three copies of $15$, each transforming in the adjoint representation of an $SU(4)$ in $SU(4)^3$.  

A similar analysis can be performed at higher orders of $t$.  Moreover, the unrefined Hilbert series, \ie~ all $x_i$, $y_i$, $z_i$ are set to $1$, can be computed from \eref{T4inD4}:
\bea
 H(t; \vec 1, \vec 1, \vec 1) &= 1 + 45 t + 128 t^{3/2}+1249 t^2 + 5504 t^{5/2}+\ldots~;
\eea
the result is in agreement with \cite{Gadde:2011uv}.

\subsubsection{$D_4$ punctures: $(5,3)$, $(5,3)$, $(2^4)$ and $(3,1^5)$}
 The quiver diagram of the mirror of this Sicilian theory can be obtained by gluing the quiver diagrams of $T_{(5,3)} (SO(8))$, $T_{(5,3)} (SO(8))$, $T_{(2^4)} (SO(8))$ and $T_{(3,1^5)} (SO(8))$ via the common symmetry $SO(8)/\bZ_2$; this is depicted in \eref{quivex3}, where each gray node labeled by $N$ denotes an $SO(N)$ gauge group (with the central node $8*$ being $SO(8)/\BZ_2$) and each black node labeled by $M$ denotes a $USp(M)$ gauge group.
{\large
\bea \label{quivex3}
\Node{}{2}-\node{}{4}-\Node{}{6}-\node{\Ver{}{2}}{\Uer{}{2}}{8*}-\Node{}{4}-\node{}{4}-\Node{}{2} -\node{}{2}  \ \tikz[na]\node(B1){};
\eea}
The quaternionic dimension of the Coulomb branch of this theory, equal to the sum of the the ranks of all gauge groups, is $18$. 

The global symmetries associated with each puncture are respectively trivial, trivial, $USp(4)$ and $SO(5) \simeq USp(4)$.  According to the top diagram of page 32 of \cite{Chacaltana:2011ze}, this Sicilian theory can be identified with the $G_2$ gauge theory with 4 fundamental hypermultiplets and 4 free hypermultiplets, which has $USp(8)$ flavor symmetry.  Indeed, the quaternionic dimension of the Higgs branch of this theory is equal to $\frac{1}{2}(7 \times 4)+4 = 18$; this is in agreement with the dimension of the Coulomb branch of the mirror theory.

\subsubsection*{The Higgs branch Hilbert series of $G_2$ gauge theory with 4 flavors of fundamental hypers, plus 4 free hypers}
In the following, we write
\bea
\tau = t^{1/2}~.
\eea
The $F$-flat Hilbert series is given by
\bea
\fflat (\tau; \vec z; \vec x) = \PE \left[\tau \chi^{USp(8)}_{[1,0,0,0]} (\vec x) \right] \times \PE \left[ \tau \chi^{USp(8)}_{[1,0,0,0]} (\vec x)  \chi^{G_2}_{[1,0]} (\vec z) -\tau^2 \chi^{G_2}_{[0,1]} (\vec z) \right]~.
\eea
The Higgs branch Hilbert series can be obtained by integrating over the $G_2$ gauge group as follows:
\bea
g(\tau, \vec x) = \int {\rm d} \mu_{G_2} (\vec z) ~\fflat (\tau; \vec z; \vec x) ~,
\eea
where the Haar measure of $G_2$ is given by
\be
\begin{split}
\int {\rm d} \mu_{G_2} (\vec z) &= \frac{1}{(2 \pi i)^2} \oint_{|z_1| =1} \frac{{\rm d} z_1}{z_1}\oint_{|z_2| =1} \frac{{\rm d} z_2}{z_2} (1-z_1)(1-z_1^2 z_2^{-1}) (1-z_1^3 z_2^{-1})  \\
& \qquad (1-z_2)(1-z_2 z_1^{-1})(1-z_2^2 z_1^{-3})~.
\end{split}
\ee
The first few terms in the power series of the Higgs branch Hilbert series $g(\tau, \vec x)$ are
\bea \label{G24flv4free}
g(\tau, \vec x) &= \PE \left[\tau \chi^{C_4}_{[1,0,0,0]} (\vec x) \right] \times \Big\{ 1+ \chi^{C_4}_{[2,0,0,0]} (\vec x) \tau^2 + \left(\chi^{C_4}_{[1,0,0,0]} (\vec x) + \chi^{C_4}_{[0,0,1,0]} (\vec x) \right) \tau^3  \nn \\
& \quad + \left(\chi^{C_4}_{[4,0,0,0]} (\vec x) + \chi^{C_4}_{[0,1,0,0]} (\vec x)+ \chi^{C_4}_{[0,2,0,0]} (\vec x)+ \chi^{C_4}_{[0,0,0,1]} (\vec x) +1 \right) \tau^4 +\ldots  \Big \}~.
\eea
Below we reproduce this Hilbert series from the Coulomb branch of the mirror theory of this Sicilian theory.  
\subsubsection*{The Coulomb branch Hilbert series of the mirror theory}
The Coulomb branch Hilbert series is given by
\be
\begin{split}
 H(t; \vec x, \vec y) &= \sum_{a_1, a_2, a_3,a_4 \geq 0} t^{-\delta_{SO(8)}(\vec n(\vec a))} P_{SO(8)} (t;\vec n(\vec a)) \; H[T_{(5,3)} (SO(8))] (t; \vec n(\vec a))\times \\
& \qquad H[T_{(5,3)} (SO(8))] (t; \vec n(\vec a)) H[T_{(2^4)} (SO(8))] (t; \vec y; \vec n(\vec a)) \times \\
& \qquad H[T_{(3, 1^5)} (SO(8))] (t; \vec x; \vec n(\vec a))~,
\end{split}
\ee
where $\vec x= (x_1, x_2)$ and $\vec y = (y_1, y_2)$ are fugacities for $SO(5)$ and $USp(4)$ respectively, and
\be
\begin{split}
H[T_{(3, 1^5)} (SO(8))] (t; \vec x; \vec n)&= t^{\frac{1}{2}\delta_{SO(8)} (\vec n)} (1-t)^4 K_{(3, 1^5)} (\vec x;t) \Psi_{D_4} (1, t , x_1, x_2;\vec n; t)~,\\
K_{(3, 1^5)} (\vec x;t) &= \PE \left[ t \chi^{B_2}_{[0,2]}(\vec x) + t^2 (1+ \chi^{B_2}_{[1,0]}(\vec x) \right]~.
\end{split}
\ee
We have checked that the first few terms in the power series of this Hilbert series agree with \eref{G24flv4free}.  In particular, the unrefined Hilbert series is
\bea
H[T_{(3, 1^5)} (SO(8))] (t, \vec x = \vec 1, \vec y =1) & =1 + 8 t^{1/2} + 72 t + 464 t^{3/2} + 2782 t^2+\ldots  \nn\\
&= \frac{1}{(1-t^{1/2})^8} (1 + 36 t + 56 t^{3/2} + 708 t^2+ \ldots)~.
\eea

\section{Coulomb branch Hilbert series of $3d$ theories with tri-vertices} \label{sec:triv}
In this section we consider the Coulomb branch of theories on {\it two} M5-branes compactified on a Riemann surface with punctures times a circle of vanishing size.  The latter are referred to as $3d$ $SU(2)$ Sicilian theories \cite{Benini:2009mz, Benini:2010uu}, or $3d$ theories with tri-vertices \cite{Hanany:2010qu}.  We emphasize that in this section we aim to compute the Coulomb branch Hilbert series of tri-vertex theories, in contrast to section \ref{sec:generalgande} in which we considered the Coulomb branch of their mirrors.

We follow the notation adopted in \cite{Hanany:2010qu}. The Lagrangian of a tri-vertex theory is specified by a graph made of tri-valent vertices connected by lines.  Each line denotes an $SU(2)$ group; an internal line (of finite length) denotes a gauge group, whereas an external line (of infinite length) denotes a flavor group. Each vertex denotes $8$ half-hypermultiplets in the tri-fundamental representation of the corresponding $SU(2)^3$ group. 
Such graphs are classified topologically by the genus $g$ and the number $e$ of external legs.  It was found in \cite{Hanany:2010qu} that the Higgs branch of such theories depends only on $g$ and $e$ and not on the details of how the vertices are connected to each other.

In this section we focus only on the cases with $g=0$, \ie~ tree diagrams, since for higher genus the theory is bad. For $g=0$, the Coulomb branch Hilbert series can be evaluated explicitly and it depends only on the number of external legs $e$ and {\it not} on the details of the graph. In section \ref{sec:genfunctrivertex} we present certain generating functions and recursive formulae that serve as powerful tools for computing Hilbert series of these class of theories using gluing techniques.  The fact that such generating functions depend solely on the number of external legs $e$ is proven in section \ref{sec:symmetry}.


It would be interesting to understand how to compute the Coulomb branch Hilbert series of theories with higher genus by determining whether they flow to a good theory in the IR.

\subsection{The case of $g=0$} \label{sec:g0}
We consider the Coulomb branch of $3d$ $\cN=4$ gauge theories based on tri-vertex tree ($g=0$) diagrams, with $SU(2)$ gauge groups associated to internal edges and tri-fundamental half-hypermultiplets associated to nodes. 

We will see in a few examples that, as for the Higgs branch \cite{Hanany:2010qu}, the Coulomb branch only depends on the number of external edges; see an example for $g=0$ and $e=6$ in section \ref{sec:g0e6} below.  We give a general proof of this fact in subsection \ref{sec:symmetry}.

\subsubsection{General formula for $g=0$ and any $e$}
\begin{figure}[H]
\begin{center}
\includegraphics[scale=0.5]{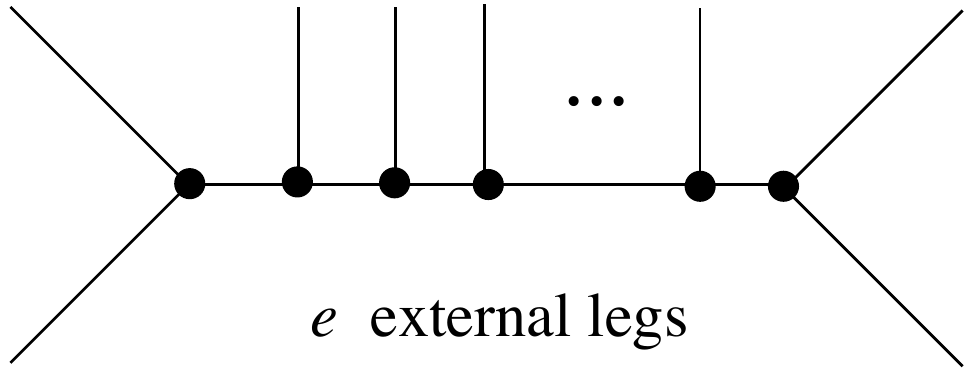}
\caption{A tri-vertex diagram with genus zero and $e$ external legs. The number of gauge groups is $e-3$.}
\label{fig:g0e}
\end{center}
\end{figure}

In the following we restrict to linear diagrams where each tri-vertex has one external leg, except for those at the ends of the line which have two external legs; see \fref{fig:g0e}.  

Let us consider $e=n+3$ external legs. The gauge group is $SU(2)^n$.
The Hilbert series of the Coulomb branch of this gauge theory is 
\be\label{HS_Coulomb_trivertices_tree}
H[g=0, e=n+3](t)= \sum_{a_1=0}^\infty \dots \sum_{a_n=0}^\infty t^{\Delta(a)} \prod_{i=1}^n P_{SU(2)}(t;a_i)\;.
\ee
The dimension formula for monopole operators is 
\be\label{mondim_trivertices_tree}
\begin{split}
\Delta(\vec a)
&= \frac{1}{2} \left[  2(|a_1|+ \left|-a_1 \right|)+{ \sum_{s_{1,2} = 0}^1 \sum_{j=1}^{n-1}  \left| (-1)^{s_1} a_j + (-1)^{s_2} a_{j+1} \right | }+2(|a_n|+\left| -a_n \right|)  \right]  \\
&- \sum_{i=1}^n \left| 2 a_i \right| =-2\sum_{i=2}^{n-1} |a_i| + \sum_{i=1}^{n-1}(|a_i-a_{i+1}|+|a_i+a_{i+1}|)\;,
\end{split}
\ee
where $a_i$, $i=1,\dots,n$ are the GNO charges in the weight lattice of the GNO dual $SO(3)^n$ group: $a_i\in\bZ_{\geq 0}$. The classical factor accounts for the Casimir invariants of the residual gauge group which is not broken by the monopole flux. For an $SU(2)$ gauge group, the classical factor is 
\be\label{classical_SU(2)}
P_{SU(2)}(t;a)= \begin{cases}
\frac{1}{1-t^2}\;, &\quad a=0 \\
\frac{1}{1-t}\;, &\quad a>0 
\end{cases}\;.
\ee

The result for the Hilbert series \eqref{HS_Coulomb_trivertices_tree} appears to be 
\be\label{HS_Coulomb_trivertices_tree_result}
\begin{split}
&H[g=0,\,e=n+3](t) = \frac{\sum_{j=0}^n \binom{n}{j} \left[\binom{n}{j} t^{2 j} - \binom{n}{j + 1} t^{2 j + 1}\right]}{(1 - t)^{2n} (1 + t)^n} \\
&= \frac{1}{(1 - t)^{2n} (1+t)^n} \Big[ -nt \; {}_2 F_1 (1-n, -n ; 2; t^2) +{}_2 F_1 (-n, -n ; 2; t^2) \Big]~.
\end{split}
\ee

\subsubsection{Special case of $g=0$ and $e=6$} \label{sec:g0e6}
\begin{figure}[H]
\begin{center}
\includegraphics[scale=0.5]{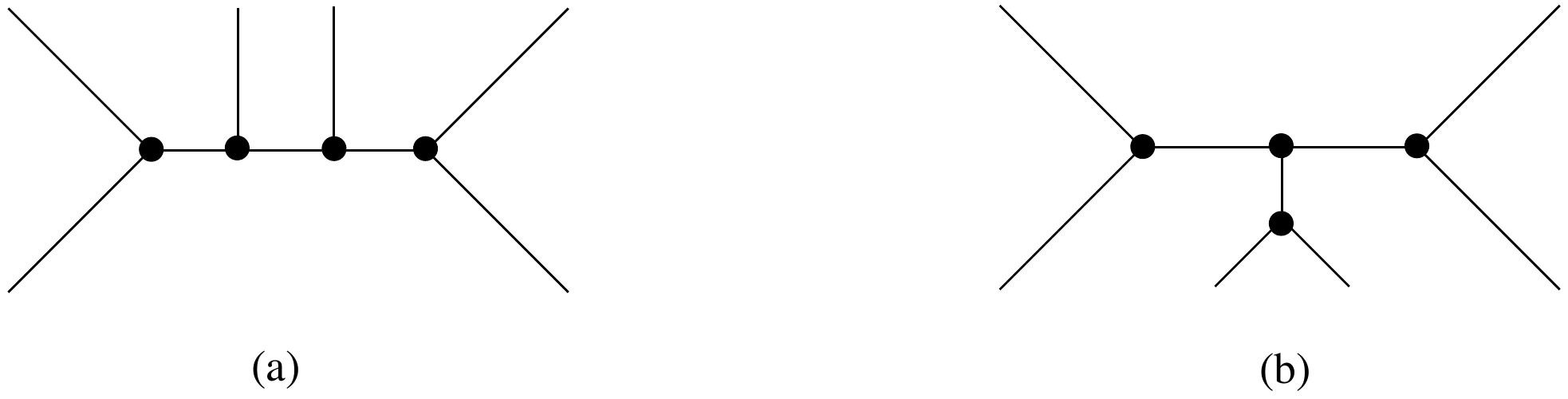}
\caption{Two tri-vertex diagrams with genus zero and $6$ external legs.}
\label{fig:g0e6}
\end{center}
\end{figure}

As an example of the fact that the Coulomb branch only depends on the number of external edges we  consider the case $e=6$.
There are two diagrams corresponding to $g=0$ and $e=6$, depicted in \fref{fig:g0e6}.
\paragraph{Diagram (a).} The Coulomb branch Hilbert series of diagram (a) is given by \eref{HS_Coulomb_trivertices_tree_result}:
\bea
H_{(a)}(t)= \frac{1-3 t+9 t^2-9 t^3+9 t^4-3 t^5+t^6}{(1-t)^6 (1+t)^3}
\eea
\paragraph{Diagram (b).}  For diagram (b), we have
\bea
\Delta_{(b)}(\vec a)
&= \frac{1}{2} \left[ \frac{1}{2} \sum_{s_{1,2,3} =0}^1\left| \sum_{i=1}^3 (-1)^{s_i} a_i \right|+2 \sum_{i=1}^3 \left( |a_i|+\left| -a_i \right| \right) \right] - \sum_{i=1}^3 |2a_i|~,
\eea
Observe that this is not equal to $\Delta_{(a)}(\vec a)$ which is given in \eref{mondim_trivertices_tree}.
However, the Hilbert series of the Coulomb branch is given by
\be
\begin{split}
H_{(b)}(t)&= \sum_{a_1,a_2,a_3=0}^\infty t^{\Delta_{(b)}(\vec a)} \prod_{i=1}^3 P_{SU(2)}(t; a_i) \\
&= \frac{1-3 t+9 t^2-9 t^3+9 t^4-3 t^5+t^6}{(1-t)^6 (1+t)^3} = H_{(a)}(t)~,
\end{split}
\ee
which is indeed equal to that of diagram (a).

\subsection{Turning on background fluxes}
So far we have computed the Coulomb Hilbert series without considering the background monopole charges coming from the global symmetries of the theory.   In this section, we turn on such background charges for the flavor symmetries present in the theory and the corresponding Hilbert series will, of course, depend on such charges.  This will turn out to be extremely useful in subsequent computations. 

Let us first consider the $T_2$ theory ($g=0, \; e=3$).   The Coulomb branch Hilbert series with background fluxes turned on is simply
\bea
H[T_2] (t; a_1, a_2, a_3) = t^{\Delta_{g=0, e=3} (a_1, a_2, a_3)}~, \label{HT2aaa}
\eea 
where $a_1, a_2, a_3 \geq 0$ are the background fluxes and
\bea
\Delta_{g=0, e=3} (a_1, a_2, a_3) = \frac{1}{2} \left[ \frac{1}{2}\sum_{s_{1,2,3}=0}^1 \left | (-1)^{s_1} a_1+ (-1)^{s_2} a_2 + (-1)^{s_3} a_3 \right | \right]~.
\eea

The Coulomb branch Hilbert series with background fluxes turned on can be handled more easily if we introduce extra fugacities to keep track of such background charges.  In this way, we end up computing the generating function of the Coulomb branch Hilbert series.  This is the topic of the next section.

\subsection{Generating functions of Coulomb branch Hilbert series} \label{sec:genfunctrivertex}
For a theory with genus zero and $e$ external legs, we can construct a generating function 
\bea
G_{e} (z_1, \ldots, z_e) = \sum_{a_1=0}^\infty \cdots \sum_{a_e =0}^\infty  H[{e}](t; a_1, \ldots, a_e) \prod_{i=1}^e z_i^{a_i} ~,
\eea
where $a_1, \ldots, a_e$ are the background fluxes for the $SU(2)^e$ global symmetry group associated to the external legs and $H[{e}](t; a_1, \ldots, a_e)$ is the usual Hilbert series with these background fluxes turned on.  Note that we omit the $t$ dependence in $G_{e} (z_1, \ldots, z_e)$ for the sake of brevity.
  To turn off the background fluxes, we simply set all $z_i$ to zero:
\bea
H[e](t; 0, \ldots, 0) = G_{e} (0,0, \ldots,0)~.
\eea

We go over the computations of generating functions in the examples below.

\paragraph{The $T_2$ theory.} From \eref{HT2aaa}, we have
\bea
G_{ e=3}(z_1, z_2, z_3) = \sum_{a_1=0}^\infty\sum_{a_2=0}^\infty\sum_{a_3=0}^\infty  t^{\Delta_{e=3} ( \vec a)} z_1^{a_1} z_2^{a_2} z_3^{a_3}~,
\eea
Evaluating the summations, we obtain
\bea \label{G03}
G_{ e=3}(\vec z) & = \frac{1}{\prod_{i=1}^3 (1-t^2 z_i) \prod_{ 1\leq j< k \leq 3} (1-t^2 z_j z_k)} \times \Big[ 1+z_1 z_2 z_3 t^3 \nn \\
& \quad +\left(-z_1 z_2-z_1 z_3-z_2 z_3-3 z_1 z_2 z_3\right) t^4+\left(-z_1^2 z_2 z_3-z_1 z_2^2 z_3-z_1 z_2 z_3^2\right) t^5 \nn \\
& \quad +2 \left(z_1 z_2 z_3+z_1^2 z_2 z_3+z_1 z_2^2 z_3+z_1 z_2 z_3^2\right) t^6+\left(z_1^2 z_2^2 z_3+z_1^2 z_2 z_3^2+z_1 z_2^2 z_3^2\right) t^7 \nn \\
& \quad +\left(-z_1^2 z_2^2 z_3-z_1^2 z_2 z_3^2-z_1 z_2^2 z_3^2\right) t^8-z_1^2 z_2^2 z_3^2 t^9 \Big]~.
\eea
Observe that $G_{ e=3}(\vec z)$ is invariant under the permutations of $z_1, z_2, z_3$.
Upon setting $z_1=z_2=z_3=0$, we recover the (trivial) Hilbert series of the Coulomb branch as expected:
\bea
G_{e=3} (0,0,0)=1~.
\eea
%

\subsubsection{Gluing generating functions and recursive formula}
If we glue a tree diagram with $e_1$ external legs with another tree diagram with $e_2$ external legs  via an external leg, the resulting diagram is a tree diagram of $e_1+e_2-2$ external legs.  In terms of the Coulomb branch Hilbert series, this gluing operation can be formulated as
\be
\begin{split}
& H[e_1+e_2-2] (\vec a)   \\
&= \sum_{a=0}^\infty H[e_1] (a_1, \ldots, a_{e_1-1}, a) P_{SU(2)} (t; a) t^{-2a} H[e_2] (a,a_{e_1}, \ldots, a_{e_1+e_2-2})~,
\end{split}
\ee
where in this formula we glue the $e_1$-th external leg of the first diagram with the first leg of the second diagram.  In terms of the generating functions, we have
\be \label{gluegenerating}
\begin{split}
G_{e_1+e_2-2} (\vec a) & = \oint_{|u|=1} \frac{\mathrm{d} u}{2 \pi i u} \oint_{|w|=1} \frac{{\rm d} u}{2 \pi i w}  \sum_{a=0}^\infty G_{e_1} (z_1, \ldots, z_{e_1-1}, u) \times \\
& \qquad u^{-a} P_{SU(2)} (t;a) t^{-2a} w^{-a}  G_{e_2} (w,z_{e_1}, \ldots, z_{e_1+e_2-2})~,
\end{split}
\ee

\subsubsection*{The recursive formula}
The diagrams with $g=0$ and $e+1$ external legs can be constructed recursively by gluing the diagram $(g=0, e=3)$ with another diagram with $e$ external legs.  We can thus obtain the recursive formula for the generating functions as follows.

From \eref{gluegenerating} we obtain
\be
\begin{split}
G_{e+1} (\vec z) & = \oint_{|u|=1} \frac{\mathrm{d} u}{2 \pi i u} \oint_{|w|=1} \frac{{\rm d} w}{2 \pi i w}  \sum_{a=0}^\infty G_{e=3} (z_1, z_2, u) \times  \\
& \qquad u^{-a} P_{SU(2)} (t; a) t^{-2a} w^{-a}  G_{e} (w,z_{3}, \ldots, z_{e+1})~.
\end{split}
\ee
We write the infinite sum as follows:
\bea
\sum_{a=0}^\infty  u^{-a} P_{SU(2)} (t; a) t^{-2a} w^{-a} = -\frac{t}{1-t^2} + \frac{t^2 uw }{(1-t)(t^2 u w-1)}~. 
\eea
Thus we have
\be
\begin{split}
G_{e+1} (\vec z) &= -\frac{t}{{1-t^2}} \oint_{|u|=1} \frac{\mathrm{d} u}{2 \pi i u} \oint_{|w|=1} \frac{{\rm d} w}{2 \pi i w}G_{e=3} (z_1, z_2, u)G_{e} (w,z_{3}, \ldots, z_{e+1}) \\
& \quad + \frac{t^2}{1-t}  \oint_{|u|=1} \frac{\mathrm{d} u}{2 \pi i } \oint_{|w|=1} \frac{{\rm d} w}{2 \pi i } \frac{G_{e=3} (z_1, z_2, u)G_{e} (w,z_{3}, \ldots, z_{e+1}) }{t^2 u w -1} \\
&= -\frac{t}{{1-t^2}} G_{e=3} (z_1, z_2, 0)G_{e} (0,z_{3}, \ldots, z_{e+1}) +  \\
& \quad + \frac{1}{1-t} \oint_{|w|=1}  \frac{\mathrm{d} w}{2 \pi i }  \frac{G_{e=3} (z_1, z_2, w^{-1}t^{-2})G_{e} (w,z_{3}, \ldots, z_{e+1}) }{w}~.
\end{split}
\ee
In the integral of the last line, we see from \eref{G03} that the poles of $G_{e=3} (z_1, z_2, w^{-1}t^{-2})$ are at $w= 1$, $w= z_1$ and $w=z_2$.  Using the residue theorem, we obtain
\be \label{recur1}
\begin{split}
G_{ e+1} (\vec z) &= -\frac{t}{{1-t^2}} G_{e=3} (z_1, z_2, 0)G_{e} (0,z_{3}, \ldots, z_{e+1}) + \\
& \quad + \frac{1}{(1-t)} {\rm Res} \Big[ G_{e=3} (z_1, z_2, w^{-1}t^{-2})  ;w =1\Big] G_{e} (1,z_{3}, \ldots, z_{e+1})\\
& \quad + \frac{1}{(1-t)z_1}  {\rm Res} \Big[ G_{e=3} (z_1, z_2, w^{-1}t^{-2})  ;w =z_1\Big] G_{e} (z_1,z_{3}, \ldots, z_{e+1}) \\
& \quad + \frac{1}{(1-t)z_2}  {\rm Res} \Big[ G_{e=3} (z_1, z_2, w^{-1}t^{-2})  ;w =z_2\Big] G_{e} (z_2,z_{3}, \ldots, z_{e+1})~.
\end{split}
\ee
Using \eref{G03}, we find that the above residues can be written in terms of simple rational functions:
\bea
{\rm Res}\Big[ G_{e=3} (z_1, z_2, w^{-1}t^{-2})  ;w =1\Big]  &= \frac{1}{\left(1-z_1\right) \left(1-z_2\right)}~, \nn \\
{\rm Res}\Big[ G_{e=3} (z_1, z_2, w^{-1}t^{-2})  ;w =z_1\Big]  &= \frac{(1-t) z_1^2 \left(-z_1-t z_1+t z_2-t z_1 z_2+t^2 z_1 z_2+t^3 z_1^2 z_2\right)}{ \left(z_1-z_2\right)  \left(1-z_1\right) \left(1-t^2 z_1\right)\left(1-t^2 z_1 z_2\right)}~, \nn \\
{\rm Res}\Big[ G_{e=3} (z_1, z_2, w^{-1}t^{-2})  ;w =z_1\Big]  &= \frac{(1-t) z_2^2 \left(-z_2+t z_1-t z_2-t z_1 z_2+t^2 z_1 z_2+t^3 z_1 z_2^2\right)}{\left(z_2-z_1\right) \left(1-z_2\right) \left(1-t^2 z_2\right) \left(1-t^2 z_1 z_2\right)}~.
\eea
We can thus rewrite \eref{recur1} as
\be\label{recur2}
\begin{split}
G_{ e+1} (\vec z) &= -\frac{t(1-t^4 z_1 z_2)}{(1-t^2)\left(1-t^2 z_1\right) \left(1-t^2 z_2\right) \left(1-t^2 z_1 z_2\right)} G_{e} (0,z_{3}, \ldots, z_{e+1}) \\
& \quad + \frac{1}{(1-t) \left(1-z_1\right) \left(1-z_2\right)}G_{e} (1,z_{3}, \ldots, z_{e+1}) \\
& \quad + \Big \{ \frac{z_1 \left(-z_1-t z_1+t z_2-t z_1 z_2+t^2 z_1 z_2+t^3 z_1^2 z_2\right)}{ \left(z_1-z_2\right)  \left(1-z_1\right) \left(1-t^2 z_1\right)\left(1-t^2 z_1 z_2\right)} G_{e} (z_1,z_{3}, \ldots, z_{e+1})  \\
& \qquad \quad  +(z_1 \leftrightarrow z_2) \Big \}~.
\end{split}
\ee

\paragraph{The special case of $z_2=\cdots = z_{e+1} =0$.}  In this case, let us denote
\bea
\widehat{G}_{e}(z) := G_{e} (z,0, \ldots,0)~.
\eea
It is immediate from \eref{recur2} that
\be \label{recursimp}
\begin{split}
\widehat{G}_{e+1}(z) &= -\frac{t}{(1-t^2)(1-t^2 z)} \widehat{G}_{e}(0)+ \frac{1}{(1-t) \left(1-z\right) }  \widehat{G}_{e}(1) \\
& \quad -\frac{(1+t) z}{(1-z) \left(1-t^2 z\right)}\widehat{G}_{e}(z)~.
\end{split}
\ee
The ordinary Hilbert series without background fluxes is obtained from $\widehat{G}_{e}(z)$ by setting $z=0$:
\bea
H[e] (t)= \widehat{G}_{e}(0)
\eea
Hence one can use the recurrence relation \eref{recursimp} to check the exact result \eref{HS_Coulomb_trivertices_tree_result}.

\subsubsection{Proof of the symmetry of the generating functions $G_{ e} (\vec z)$}\label{sec:symmetry}

The Coulomb branch Hilbert series of the tri-vertex theories only depends on the number of external legs. This follows from the fact that $G_{e}(z_1,\cdots ,z_e)$
is a symmetric function of the variables $z_1,\cdots,z_e$.  In this section we sketch a proof of this statement. 

The proof  goes trough various steps.

\ben
\item  We have seen in (\ref{G03}) that $G_{e=3}(z_1,z_2,z_3)$ is invariant under permutations of $z_1,z_2,z_3$. Using (\ref{G03}) and the recursion relation (\ref{recur2})
we can evaluate $G_{e=4}(z_1,z_2,z_3,z_4)$, whose expression is too long to be reported here, and explicitly check that it  is invariant under permutations of $z_1,z_2,z_3,z_4$.

\item We next analyze {\it linear}  tree-level theories  consisting of a linear chain of $e-2$ vertices, each connected to the following one by an internal line, and with a total number of  $e$ external  legs.  An example for the case $e=6$ is given in part (a) of figure \ref{fig:g0e6}. We now show that the  generating function $G_{e}(z_1,\cdots ,z_e)$ for a {\it linear} theory   is fully symmetric in the $z_i$. It is enough to show that it is invariant under the exchange of any pair of neighboring external legs. Let $z_i$ and $z_{i+1}$ the fugacities
associated with the pair of external legs. We can always obtain the {\it linear}  theory by gluing a $e=4$ tree diagram containing the two external legs $z_i$ and $z_{i+1}$ with two {\it linear} theories with $i$ and $e-i$ external legs and write
\bea
 G_{e}(z_1,\cdots ,z_e) =\sum_{a,a^\prime=0}^\infty & G_{i}(z_1,\cdots ,z_{i-1}, a) P_{SU(2)} (t; a) t^{-2a} 
G_{e=4}(a,z_i,z_{i+1},a^\prime )&\nonumber\\  &  P_{SU(2)} (t; a^\prime) t^{-2a^\prime} G_{e-i}(a^\prime, z_{i+2},\cdots ,z_{e}) \, ,&
\eea
where the symmetry in $z_i$ and $z_{i+1}$ is manifest.

\item   A generic genus zero tri-vertex theory also contains {\it saturated} vertices, i.e. vertices that are connected to three other different vertices by internal lines.  We now show that any genus zero diagram can be reduced to a {\it linear} one with the same generating function. This will prove our statement for all genus zero theories. As an example  we can consider the theory in part (b) of figure \ref{fig:g0e6}. We can recognize that the diagram is obtained by gluing two simple three-vertices $(g=0,e=3)$ with a four-vertex diagram $(g=0,e=4)$, and its generating function can be written as 
\be\label{glue334}
\begin{split} 
G_{e=6}(z_1,\cdots ,z_6) =\sum_{a,a^\prime=0}^\infty & G_{e=3}(z_1,z_2, a) P_{SU(2)} (t; a) t^{-2a} 
G_{e=4}(a, a^\prime , z_3,z_4)  \\  &  P_{SU(2)} (t; a^\prime) t^{-2a^\prime} G_{e=3}(a^\prime, z_5,z_6) \, .
\end{split}
\ee
Since the four-vertex diagram is fully symmetric under the exchange of the external legs, we can permute them and give a different shape to our diagram.
In particular, equation (\ref{glue334}) is also the generating function for the {\it linear} diagram  in part (a) of figure \ref{fig:g0e6}. In a similar way, whenever
a {\it linear} diagram is attached to a {\it saturated} node by gluing the two external legs at one of its extremities, by permuting its legs 
we can remove the   {\it saturated} node in favor of a linear structure. By repeating this process many times we can transform any genus zero diagram into a {\it linear} one.


\een

This ends our proof. We notice that we can construct the Hilbert series of higher genus tri-vertex theories  by identifying external legs of a genus zero graph, adding the appropriate factor  $P_{SU(2)} (t,a)$, the contribution of the gauge fields to the dimension formula  and summing over the $a$. Unfortunately, since the resulting theory is bad, the Hilbert series is divergent. We can make it finite by changing the matter content
and adding matter fields transforming under the gauge groups of the legs that are identified. For example, by adding one or more adjoint hypermultiplets to each leg that has been identified the Hilbert series becomes
convergent. As a curiosity, we notice that the resulting Hilbert series will be fully symmetric under the exchange of the external legs, since the tree-level Hilbert series was. It would be interesting to see if any of these
regularized theories are related to the IR behavior of the higher genus tri-vertex theories.

\section{Conclusion}

In this paper we have applied gluing techniques to the computation of the Coulomb branch Hilbert series of mirrors of three dimensional Sicilian theories  and we have successfully compared our results with the superconformal index predictions for the Higgs branch of the Sicilian theories themselves. As shown in \cite{Gadde:2011uv}, the Hall-Littlewood limit of the $4d$ $\cN=2$ superconformal index captures the Higgs branch Hilbert series only for genus zero Riemann surfaces. One of the main results of this paper is formula \eqref{genus0} for genus zero: it perfectly agrees with the findings of \cite{Gadde:2011uv},  that were obtained in a completely different manner. 
  
We have also computed the Coulomb branch Hilbert series of mirrors of Sicilian theories with genus greater than one. For $N=2$ M5-branes, the Sicilian theories are Lagrangian and their Higgs branch Hilbert series can be computed by standard methods \cite{Hanany:2010qu}. We have successfully matched those results with our Coulomb branch Hilbert series of the mirror theories, providing a check of our formulas based on mirror symmetry. For $N>2$, there is no other available method for computing the Higgs branch Hilbert series of Sicilian theories. Our results give non-trivial predictions, that would be nice to check in some other way, maybe using the $3d$ superconformal index.

Our results clarify why the Hall-Littlewood polynomials appear in two different contexts,  the Coulomb branch Hilbert series for the $T_{\vec \rho}(G)$ theories and the four dimensional superconformal index of Sicilian theories. It is interesting to see how the Hall-Littlewood limit of the superconformal index formula \cite{Gadde:2011uv,Lemos:2012ph}, 
emerging from an apparently unrelated construction, can be naturally reinterpreted in terms of gluing of three dimensional building blocks. It would be interesting to see if these Hilbert series are computed by some auxiliary three or two dimensional topological theories along the lines of \cite{Gadde:2009kb,Gadde:2011ik,Moore:2011ee}.

It is natural to expect that the gluing prescription discussed in this paper can be generalized to any group $G$, including non-simply laced and exceptional groups, by gluing Coulomb branch Hilbert series \eref{mainHL} of $T_{\vec\rho}(G)$ tails via the common centerless symmetry $G/Z(G)$ in the obvious way.  This should yield the Coulomb branch Hilbert series for the mirror of 5d $\cN=2$ super Yang-Mills of gauge group $G$ compactified on the punctured Riemann surface. It would be nice to come up with an explicit check for this proposal.

Our results clearly show that gluing is an efficient technique to evaluate the Coulomb branch Hilbert series, once the Hilbert series with background fluxes of the building blocks are explicitly known. It would be interesting to extend our analysis to cover  more general classes of building blocks which can be applied to an  even wider class of $\cN = 4$ gauge theories.


\section*{Acknowledgements}

We thank Francesco Benini, Nick Halmagyi, Yuji Tachikawa and Alessandro Tomasiello for useful discussions, and the following institutes and workshops for hospitality and partial support: the Galileo Galilei Institute for Theoretical Physics and INFN and the Geometry of Strings and Fields workshop (SC), the Simons Center for Geometry and Physics and the 2013 Summer Workshop, and Chulalongkorn University and the 3rd Bangkok Workshop on High Energy Theory (AH and NM), \'Ecole Polytechnique and the String Theory Groups of the universities of Rome ``Tor Vergata'' and of Oviedo (NM). NM is also grateful to Diego Rodr\'iguez-G\'omez, Yolanda Lozano, Raffaele Savelli, Jasmina Selmic, Hagen Triendl, Sarah Maupeu and Mario Pelliccioni for their very kind hospitality. We were partially supported by the STFC Consolidated Grant ST/J000353/1 (SC), the EPSRC programme grant EP/K034456/1 (AH), the ERC grant, Short Term Scientific Mission of COST Action MP1210, and World Premier International Research Center Initiative (WPI Initiative), MEXT, Japan (NM), and  INFN and the MIUR-FIRB grant RBFR10QS5J ``String Theory and Fundamental Interactions'' (AZ).

%


\appendix
\section{Mirrors of Sicilian theories with twisted $D$ punctures}\label{app:twisted}
In this appendix, we briefly discuss $3d$ Sicilian theories with twisted $D$ punctures.  A twisted $D_N$ puncture can be written in terms of a $C_{N-1}$ partition $\tilde{\vec \rho} = [ \tilde{\rho}_i]$ with $\sum_i \tilde{\rho}_i= 2N-2$ and $r_k$ the number of times that part $k$ appears. The  global symmetry to this puncture is given by
\bea
G_{\tilde{\vec \rho}} = \prod_{\text{$k$ odd}} USp(r_k) \times  \prod_{\text{$k$ even}} SO(r_k)~.
\eea
For example, the global symmetry associated with twisted $D_4$ puncture $(2, 1^4)$ is $USp(4)$.  

A building block of a mirror of Sicilian theories with twisted $D$ punctures is a gauge theory $T_{\tilde{\vec \rho}} (B_{N-1})$, whose quiver diagram, of the type first considered in \cite{Feng:2000eq}, is given by (4.3) of \cite{Cremonesi:2014kwa} and (6.5) of \cite{Benini:2010uu}. The quiver gauge theory is \emph{bad} in the sense of \cite{Gaiotto:2008ak}, therefore the monopole formula for the Coulomb branch Hilbert series diverges. However, according to \cite{Cremonesi:2014kwa}, the Coulomb branch Hilbert series of the infrared CFT is conjectured to be computed by the Hall-Littlewood formula \eref{mainHL}, which gives
\bea \label{HLtwisitedD}
H[T_{\tilde{\vec \rho}} (B_{N-1})] (t; \vec x;\vec n) = t^{\frac{1}{2} \delta_{B_{N-1}} (\vec n)} (1-t)^{N-1} K_{\vec \rho}^{C_{N-1}}(\vec x; t) \Psi^{\vec n}_{C_{N-1}}(\vec a(t, \vec x);t)~.
\eea
We will assume the validity of this formula in the rest of the appendix.

\paragraph{Example: $\tilde{\vec \rho} = (2,1^4)$.} For example, given an $SO(8)$ twisted puncture $\tilde{\vec \rho} = (2,1^4)$, the corresponding theory is
\bea \label{21111quiv}
T_{(2,1^4)} (SO(7)): \qquad [SO(7)] - (USp(4)) -(O(5)) -(USp(2))-(O(3))~.
\eea
Note that such a tail is a typical component in a mirror pair computation of \cite{Feng:2000eq}.

The Hall-Littlewood formula \eref{HLtwisitedD} applied to this theory gives the Coulomb branch Hilbert series
\bea \label{HLT21111}
H[T_{(2,1^4)} (SO(7))] (t; \vec x; \vec n)= t^{\frac{1}{2}(5n_1+3n_2+n_3)}  K_{(2,1^4)}^{USp(6)}(\vec x; t) \Psi^{(n_1,n_2,n_3)}_{USp(6)}(t, x_1, x_2;t)~,
\eea
where the notations are explained below:
\bi
\item $\vec x = (x_1, x_2)$ are the fugacities of the global symmetry $USp(4)$.
\item $\vec n =(n_1, n_2, n_3)$ are the background fluxes for $SO(7)$, with the restriction 
\bea
n_1 \geq n_2 \geq n_3 \geq 0~.
\eea
\item The argument $(t, x_1, x_2)$ of the Hall-Littlewood polynomial is obtained from the decomposition of the fundamental representation of $USp(6)$ into representations of $SU(2) \times USp(4)$:
\bea
\chi^{USp(6)}_{[1,0,0]} (\vec y) = \chi^{SU(2)}_{[1]}(t^{1/2}) + \chi^{USp(4)}_{[1,0]} (\vec x)~,
\eea
so that $\vec y =(y_1, y_2, y_3) = (t^{1/2}, x_1, x_2)$.
\item The prefactor $K_{(2,1^4)}^{USp(6)}(\vec x; t)$ comes from the following decomposition of the adjoint representation of $USp(6)$:
\bea
\chi^{USp(6)}_{[2,0,0]}(t^{1/2}, x_1, x_2) = \chi^{USp(4)}_{[2,0]} (\vec x) +  \chi^{USp(4)}_{[1,0]} (\vec x)\chi^{SU(2)}_{[1]} (t^{1/2}) +  \chi^{SU(2)}_{[2]} (t^{1/2})~.
\eea
Hence, according to \eref{K}, the prefactor is given by
\bea
K_{(2,1^4)}^{USp(6)}(\vec x; t) =  \PE \left[t \chi^{USp(4)}_{[2,0]} (\vec x) + t^{3/2}\chi^{USp(4)}_{[1,0]} (\vec x)  + t^2  \right] ~.
\eea
\ei
For reference, we provide the Hilbert series with vanishing background fluxes:
\bea
H[T_{(2,1^4)} (SO(7))] (t; \vec x; \vec 0) = \PE\left[ \chi^{USp(4)}_{[2,0]} (\vec x) t +  \chi^{USp(4)}_{[1,0]} (\vec x) t^{3/2} - t^4 -t^6 \right]~.
\eea
Hence the Coulomb branch of this theory is a complete intersection space with $12$ complex dimensions, as expected from the quiver diagram \eref{21111quiv}.

\subsection{The Coulomb branch Hilbert series of mirror theories}
Let us consider a Sicilian theory associated with a Riemann surface with genus $g$ and two sets of punctures: $2m$ twisted $D_N$ punctures $\tilde{\vec \rho}_1, \tilde{\vec \rho}_2, \ldots, \tilde{\vec \rho}_{2m}$ and $n$ untwisted $D_N$ puncture $\vec \rho_1, \vec \rho_2, \ldots, {\vec \rho}_n$. 

Following the prescription of \cite{Benini:2010uu}, the mirror of this Sicilian theory can be constructed by gluing $T_{\tilde{\vec \rho}_1} (B_{N-1}), \ldots T_{\tilde{\vec \rho}_{2m}} (B_{N-1})$ together with $T_{{\vec \rho}_1} (D_N), \ldots T_{{\vec \rho}_n} (D_N)$, with the common global symmetry group $B_{N-1}=SO(2N-1)$, which is a subgroup of $D_N=SO(2N)$, being gauged. The mirror quiver also contains $m+g-1$ hypermultiplets in the vector representation of the common gauge group $SO(2N-1)$.
We test the prescription of \cite{Benini:2010uu} in section \ref{sec:fourtwists} below using the mirror of a theory associated with a genus zero surface and four $SO(4)$ punctures.

The Coulomb branch Hilbert series of the resulting mirror theory is therefore 
\be \label{glueDtwist}
\begin{split}
&H(t, \tilde{\vec x}_1, \ldots, \tilde{\vec x}_m, \vec x_1, \ldots, \vec x_n) \\
&=\sum_{n_1 \geq \cdots \geq n_{N-1} \geq 0} t^{(g-1)\delta_{SO(2N-1)}(\vec n)} {\color{blue} t^{(m+g-1) \sum_{i=1}^{N-1} n_i}} P_{SO(2N-1)} (t;n_1, \ldots, n_{N-1}) \times  \\
& \prod_{i=1}^m H[T_{\tilde{\vec \rho}_i} (B_{N-1})] (t; \tilde{\vec x}_i; n_1, \ldots, n_{N-1})  \prod_{j=1}^n H[T_{\vec \rho_j} (D_N)] (t; \vec x_j; n_1, \ldots, n_{N-1},0)~,
\end{split}
\ee
where the Casimir factor $P_{SO(2N-1)}$ is given by (A.9) and (A.6) of \cite{Cremonesi:2013lqa} and $\delta_{SO(2N-1)}$ is given by \eref{powers}; the fugacities $\tilde{\vec x}_1, \ldots, \tilde{\vec x}_m$ correspond to the global symmetries associated with the twisted punctures $\tilde{\vec \rho}_1, \ldots, \tilde{\vec \rho}_m$ respectively, and similarly for the non-tilde fugacities.  Here the factor denoted in blue is the contribution from the extra $m+g-1$ hypermultiplets in the vector representation of the gauge group $SO(2N-1)$. 

It can be checked that formula \eref{glueDtwist} agrees with formula (4.10) of \cite{Lemos:2012ph} and formula (2.8) of \cite{Chacaltana:2013oka} for the HL index of the Sicilian theory in the case of two twisted punctures and genus 0.

Below we demonstrate formula \eref{glueDtwist} using examples with $SO(8)$ twisted and untwisted punctures on a Riemann surface with genus $0$.
\subsubsection{Twisted punctures $(2,1^4)$, $(2,1^4)$ and untwisted puncture $(4,4)$} 
Let us present an explicit example with $\tilde{\vec \rho}_1= \tilde{\vec \rho}_2=(2,1^4)$ and $\vec{\rho}_1 = (4,4)$.  The Coulomb branch Hilbert series of $T_{(4,4)}(SO(8))$ is discussed in detail in Appendix C.2 of \cite{Cremonesi:2014kwa}. From \eref{glueDtwist}, the Hilbert series of the mirror of the Sicilian theory in question is
\bea \label{SO8Dtwistex}
& H(t, \vec a, \vec b, c) \nn\\
& = \sum_{n_1 \geq n_2 \geq n_3 \geq 0} t^{-(5n_1+3n_2+n_3)} P_{SO(7)} (t; n_1,n_2,n_3) H[T_{(2,1^4)} (SO(7))] (t; a_1, a_2; n_1,n_2,n_3) \times \nn \\
& \qquad H[T_{(2,1^4)} (SO(7))] (t; b_1, b_2; n_1,n_2,n_3) H[T_{(4,4)} (SO(8))] (t; c; n_1,n_2,n_3,0)~,
\eea
where the explicit expressions for $H[T_{(2,1^4)} (SO(7))]$ and $H[T_{(4,4)} (SO(8))]$ are given by \eref{HLT21111} and by (C.18) of \cite{Cremonesi:2014kwa}, respectively.  The fugacities $\vec a =(a_1, a_2)$, $\vec b=(b_1,b_2)$ and $c$ correspond to the global symmetries $USp(4)$, $USp(4)$ and $USp(2)$ respectively.

The Higgs branch of the four dimensional Sicilian theory with the same punctures as this example was discussed on Page 30 of \cite{Chacaltana:2013oka} and Fixture 16 on Page 35 of the same reference.  Upon expanding \eref{SO8Dtwistex} in a power series in $t$, we find an agreement with \cite{Chacaltana:2013oka}, namely
\bea
H(t; \vec a, \vec b, c)  = \PE\left[ t^{1/2} \chi^{SU(2)}_{[1]} (c) \right] \times \widetilde{H}(t, \vec a, \vec b, c)~,
\eea
where the first factor with the $\PE$ denotes the free hypermultiplet whose chiral multiplets transforming as a doublet of $SU(2)$, and the first few terms in irreducible part $\widetilde{H}(t, \vec a, \vec b, c)$ are
\bea \label{refHStwistedD}
\widetilde{H}(t; \vec a, \vec b, c) &= 1+ \chi^{C_5}_{[2,0,0,0,0]} (\vec y) t + \chi^{C_5}_{[0,0,0,0,1]} (\vec y) t^{3/2} +  \nn \\
& \qquad (\chi^{C_5}_{[4,0,0,0,0]} (\vec y)+ \chi^{C_5}_{[0,2,0,0,0]} (\vec y)+1) t^{2} + \ldots~,
\eea
where $\vec y =(y_1, \ldots, y_5)$ are fugacities of $USp(10)$ and a possible fugacity map between $\vec y$ and $\vec a, \; \vec b,\; c$ is
\bea
y_1=a_1, \quad y_2=a_2, \quad y_3= c, \quad y_4= b_1, \quad y_5= b_2~.
\eea
The plethystic logarithm of \eref{refHStwistedD} indicates that there are 55 generators at order $t$ transforming in the representation $[2,0,0,0,0]$ of $USp(10)$ and 132 generators at order $t^{3/2}$ in the representation $[0,0,0,0,1]$ of $USp(10)$.

The unrefined Hilbert series $\widetilde{H}(t; \vec a =\vec 1, \vec b=\vec 1, c=1)$ can be computed exactly:
\be
\begin{split}
& \widetilde{H}(t; \vec a =\vec 1, \vec b=\vec 1, c=1) = \frac{1}{(1-t)^{32} (1+t)^{18} \left(1+t+t^2\right)^{16}} ~\times  \\
& \quad\Big( 1+2 t+40 t^2+194 t^3+1007 t^4+4704 t^5+18683 t^6+67030 t^7+220700 t^8  \\
& \quad +657352 t^9+1796735 t^{10}+4540442 t^{11}+10610604 t^{12}+23011366 t^{13} \\
& \quad+46535540 t^{14}+87887734 t^{15}+155277056 t^{16}+257288236 t^{17} \\
& \quad+400453203 t^{18}+585971786 t^{19}+807195575 t^{20}+1047954388 t^{21} \\
& \quad +1282842123 t^{22}+1481462886 t^{23}+1615002952 t^{24}+1662191888 t^{25} \\
& \quad +1615002952 t^{26} + \text{palindrome up to $t^{50}$} \Big)~.
\end{split}
\ee
The irreducible component of the Coulomb branch is 16 quaternionic dimensional, as indicated by half of order of the pole at $t=1$ in the unrefined Hilbert series.  Taking into account the free hypermultiplet, the Coulomb branch of this mirror theory is 17 quaternionic dimensional.  This agrees with the result stated in the second bullet point on Page 50 of \cite{Chacaltana:2013oka} that the difference between the effective numbers of hypermultiplets and vector multiplets is $35-18=17$.

\subsubsection{Twisted punctures $(6)$, $(1^6)$ and untwisted puncture $(1^8)$} 

From \eref{glueDtwist}, the Hilbert series of the mirror of the Sicilian theory in question is
\be \label{SO8Dtwist61618}
\begin{split}
& H(t; \vec b, \vec c) \\
& = \sum_{n_1 \geq n_2 \geq n_3 \geq 0} t^{-(5n_1+3n_2+n_3)} P_{SO(7)} (t;n_1,n_2,n_3) H[T_{(6)} (SO(7))] (t;  n_1,n_2,n_3) \times  \\
& \qquad H[T_{(1^6)} (SO(7))] (t; \vec b; n_1,n_2,n_3) H[T_{(1^8)} (SO(8))] (t; \vec c; n_1,n_2,n_3,0)~,
\end{split}
\ee
where $\vec b=(b_1,b_2,b_3)$ are fugacities for $USp(6)$ and $\vec c =(c_1, \ldots, c_4)$ are fugacities for $SO(8)$.  Here the Hilbert series of the building blocks are given by
\bea
H[T_{(6)} (SO(7))] (t;  n_1,n_2,n_3)&= t^{\frac{1}{2}(5n_1+3n_2+n_3)}  \PE[t^2+t^4+t^6] \Psi^{(n_1,n_2,n_3)}_{USp(6)}(t^{1/2},t^{3/2},t^{5/2};t) \nn \\
& = 1~, \nn \\
H[T_{(1^6)} (SO(7))] (t; \vec b;  n_1,n_2,n_3)&= t^{\frac{1}{2}(5n_1+3n_2+n_3)}  \PE[t \chi^{USp(6)}_{[2,0,0]} (\vec b)] \Psi^{(n_1,n_2,n_3)}_{USp(6)}(\vec b;t)~, \nn \\
H[T_{(1^8)} (SO(8))] (t; \vec c;  n_1,n_2,n_3,n_4)&= t^{\frac{1}{2}(6n_1+4n_2+2n_3)}  \PE[t \chi^{SO(8)}_{[0,1,0,0]} (\vec b)] \Psi^{(n_1,n_2,n_3,n_4)}_{SO(8)}(\vec c;t)~.
\eea
It can be checked that \eref{SO8Dtwist61618} is equal to the Hilbert series of $48$ free half-hypermultiplets:
\bea
H(t; \vec b, \vec c) = \PE \left[t^{1/2} \chi^{USp(6)}_{[1,0,0]} (\vec b) \chi^{SO(8)}_{[1,0,0,0]} (\vec c) \right]~.
\eea 
This confirms the free field fixture $\#8$ on Page 37 of \cite{Chacaltana:2013oka}.

\subsubsection{Four $SO(4)$ twisted punctures: $(2)$, $(2)$, $(1^2)$, $(1^2)$} \label{sec:fourtwists}
The aim of this example is to test the prescription of adding extra fundamental  hypermultiplets of $SO(2N-1)$, as discussed in \cite{Benini:2010uu}. From \eref{glueDtwist}, the Coulomb branch Hilbert series of the mirror of the $D_4$ Sicilian theory with twisted punctures $(2)$, $(2)$, $(1^2)$, $(1^2)$ is
\be
\begin{split}
& H(t; x_1, x_2) \\
& = \sum_{n \geq 0} t^{-n} {\color{blue} t^{n}} P_{SO(3)} (t;n) \Big[ H[T_{(2)} (SO(3))] (t;  n) \Big]^2  \prod_{i=1}^2 H[T_{(1,1)} (SO(3))] (t; x_i; n)  \\
& =\sum_{n \geq 0} t^{-n} {\color{blue} t^{n}} P_{SO(3)} (t; n) \prod_{i=1}^2 H[T_{(1,1)} (SO(3))] (t; x_i; n)~,
\end{split}
\ee
where the blue factor denotes the contribution of the extra hypermultiplet.  The explicit expressions of each Hilbert series in the summand are
\be
\begin{split}
H[T_{(1,1)} (SO(3))] (t; x; n) &= \frac{t^n x^{-n} \left(1-t^2 x^2+t^2 x^{2 n}-x^{2+2 n}\right)}{ (1-x^2) \left(1-t^2 x^{-2}\right)\left(1-t^2 x^2\right)}~, \\
H[T_{(2)} (SO(3))] (t;  n)&=1~.
\end{split}
\ee  

Performing the summation, we find that the above Hilbert series is equal to (D.14) of \cite{Cremonesi:2014kwa}.  Setting $x_1=x_2=1$, we recover the unrefined Hilbert series written in (D.15) of \cite{Cremonesi:2014kwa}.  This is indeed equal to the Higgs branch Hilbert series of the $SO(4)$ gauge theory with $2$ flavors, in accordance with Section 4.1 of \cite{Chacaltana:2013oka}.

\bibliographystyle{ytphys}
\bibliography{ref_v2}

\end{document}